\documentclass[twocolumn]{aastex63}
\usepackage{mathtools}
\usepackage[maxfloats=40]{morefloats} 
\usepackage{color}
\usepackage{amsmath}
\usepackage[figure,figure*]{hypcap}
\usepackage{lineno}

\DeclarePairedDelimiter\abs{\lvert}{\rvert}%
\DeclarePairedDelimiter\norm{\lVert}{\rVert}%

\makeatletter
\let\oldabs\abs
\def\abs{\@ifstar{\oldabs}{\oldabs*}}
\let\oldnorm\norm
\def\norm{\@ifstar{\oldnorm}{\oldnorm*}}
\makeatother

\newcommand{\Rom}[1]{\uppercase\expandafter{\romannumeral #1\relax}}
\def \jyb {Jy~beam$^{-1}$}
\def \mjyb {mJy~beam$^{-1}$}
\def \kms {km~s$^{-1}$}
\def\C18O{\textrm{C$^{18}$O}}
\def\nh3{\textrm{NH$_{3}$}}

\begin{document}

\title{Formation of the SDC13 Hub-Filament System: A Cloud-Cloud Collision Imprinted on The Multiscale Magnetic Field}

\accepted{2022/4/18}

\author[0000-0002-6668-974X]{Jia-Wei Wang}
\affiliation{Academia Sinica Institute of Astronomy and Astrophysics, No.1, Sec. 4, Roosevelt Road, Taipei 10617, Taiwan}
\email{jwwang@asiaa.sinica.edu.tw}

\author[0000-0003-2777-5861]{Patrick M. Koch}
\affiliation{Academia Sinica Institute of Astronomy and Astrophysics, No.1, Sec. 4, Roosevelt Road, Taipei 10617, Taiwan}

\author[0000-0002-0675-276X]{Ya-Wen Tang}
\affiliation{Academia Sinica Institute of Astronomy and Astrophysics, No.1, Sec. 4, Roosevelt Road, Taipei 10617, Taiwan}

\author[0000-0001-8509-1818]{Gary A. Fuller}
\affiliation{Jodrell Bank Centre for Astrophysics, Department of Physics and Astronomy, University of Manchester, Manchester M13 9PL, UK}
\affiliation{Physikalisches Institut, University of Cologne, Z\"ulpicher Str. 77, 50937 K\"oln, Germany}

\author[0000-0002-6893-602X]{Nicolas Peretto}
\affiliation{School of Physics and Astronomy, Cardiff University, Queens Buildings, The Parade, Cardiff CF24 3AA, UK}

\author[0000-0001-5933-2147]{Gwenllian M. Williams}
\affiliation{School of Physics and Astronomy, University of Leeds, Leeds, LS2 9JT, UK}

\author[0000-0003-1412-893X]{Hsi-Wei Yen}
\affiliation{Academia Sinica Institute of Astronomy and Astrophysics, No.1, Sec. 4, Roosevelt Road, Taipei 10617, Taiwan}

\author[0000-0002-9846-7017]{Han-Tsung Lee}
\affiliation{Academia Sinica Institute of Astronomy and Astrophysics, No.1, Sec. 4, Roosevelt Road, Taipei 10617, Taiwan}
\affiliation{Department of Physics, National Central University, Chung-Li 32054, Taiwan}
\affiliation{Institute of Astrophysics, National Taiwan University, Taipei 10617, Taiwan}

\author[0000-0002-8336-2837]{Wei-An Chen}
\affiliation{Academia Sinica Institute of Astronomy and Astrophysics, No.1, Sec. 4, Roosevelt Road, Taipei 10617, Taiwan}
\affiliation{Department of Physics, National Cheng Kung University, No.1, University Road, Tainan City 701, Taiwan}

\begin{abstract}
Hub-filament systems (HFSs) are potential sites of protocluster and massive star formation, and play a key role in mass accumulation.
We report JCMT POL-2 850 $\mu$m polarization observations toward the massive HFS SDC13. 
We detect an organized magnetic field near the hub center with a cloud-scale ``U-shape'' morphology following the western 
edge of the hub. Together with larger-scale APEX $^{13}$CO and \textit{PLANCK} polarization data, we find that SDC13 is located at the convergent point of three giant molecular clouds (GMCs) along a large-scale, partially spiral-like magnetic field. The smaller ``U-shape'' magnetic field is perpendicular to the large-scale magnetic field and the converging GMCs. We explain this as the result of a cloud-cloud collision.
Within SDC13, 
we find that local gravity and velocity gradients point toward filament ridges and hub center. 
This suggests that gas can locally be pulled 
onto filaments and overall converges to the hub center.
A virial analysis of the central hub shows that gravity dominates magnetic and kinematic energy. Combining large- and small-scale analyses, we propose that SDC13 is initially formed from a collision of clouds moving along the large-scale magnetic field. In the post-shock regions, after the initial turbulent energy has dissipated, gravity takes over and starts to drive the gas accretion along the filaments toward the hub center. 
\end{abstract}

\keywords{ISM: clouds --- ISM: magnetic fields --- ISM: structure --- ISM: individual objects (SDC13) --- ISM: kinematics and dynamics}

\section{Introduction}
Stars can form in a clustered environment \citep{la03}, and young massive clusters are born embedded within giant molecular clouds (GMCs) \citep{la03,po10}. The \textit{Herschel} Gould Belt survey shows that star-formation activities predominately take place within the density-enhanced filamentary structures of these GMCs \citep[e.g.,][]{an10,ar11,an14}. Among these filamentary structures, attention has recently been drawn to a special configuration named hub-filament system (HFS). This consists of a dense hub
towards which numerous filaments converge \citep{my09,mo18}.
Statistical analyses based on the clumps identified in the \textit{Herschel} HiGAL survey show that all massive stars and clusters preferentially form in the density-enhanced centers of HFSs \citep{ku20}. \citet{an21} find that infrared dark HFSs tend to concentrate more mass into their largest cores as compared to infrared bright hubs, and they  suggest that HFSs can efficiently concentrate mass in the early evolutionary stage. Hence, HFSs are considered a key stage to activate massive star and cluster formation. Understanding the formation and evolution of HFSs therefore plays an important role in developing our 
picture of star formation.

Molecular clouds are formed out of the atomic phase of the interstellar medium. A topic of considerable interest is how the filaments and HFSs evolve from molecular clouds, as they determine the initial conditions for star formation. The 0.1-pc width found in the diffuse, thermally subcritical filaments from the \textit{Herschel} Gould Belt Survey favors a scenario where filaments originate from large-scale compressive flows \citep{pa01a,ar11,an14}. HFSs can possibly be formed by the collision of these filamentary clouds \citep{na14,ku20}. This is supported by observations of HFS velocity structures \citep[e.g.,][]{mo19,do19,en21} and the detection of related shock tracers \citep[e.g.,][]{na14}. In addition to filament-filament collision, physical processes driven by gravitational instabilities, such as multi-scale gravitational collapse \citep{va09,go14,go18} or layer fragmentation \citep{my09,va14}, are also proposed as a possible origin of HFSs. These theories are supported by the gravity-induced patterns found in magnetic field morphologies and the filament velocity structures \citep{my09,bu13,va14,wa19,ch20,wa21}. However, the observed density and velocity structures only represent 
the current snapshot in time, 
and thus, it is difficult to constrain the forming environment of HFSs in the early evolutionary stage. In contrast to that, the larger-scale physical processes from the early evolutionary stage of molecular clouds might be imprinted and still preserved in the magnetic field morphology within the large-scale, diffuse medium. Hence, studying the variation of magnetic field morphologies at different densities and physical scales can provide insight into the evolutionary history of clouds \citep{koch12b,so13,li15,koch14,he19}.

The roles of magnetic fields in the formation of HFSs are varying, depending on the dominating physical process. In the diffuse interstellar medium, strong magnetic fields can guide large-scale MHD flows, and generate filamentary structures aligned or perpendicular to organized magnetic fields \citep[e.g.,][]{na08,pal13,li14,soam19}. Weak magnetic fields can be compressed and modified by dominating large-scale turbulence \citep[e.g.,][]{pa01a} or shocks due to expanding bubbles or cloud-cloud collisions \citep[e.g.,][]{pe12, li19}. In dense clouds, magnetic fields can be important in regulating the cloud fragmentation and collapse
\citep[e.g.,][]{na78,my09,va14,ta19,pa21}. However, magnetic fields can be shaped by gravitational collapse and gravity-induced accretion flows \citep[e.g.,][]{cr12,go18,wa19}. Generally, the detailed comparison between cloud/filament properties and magnetic field morphologies yields a useful tool to constrain the dominating physical process \citep[e.g.,][]{so13,koch12b,koch13,ko14,ko18,ta18,wa21}.

SDC13 is a well-known filamentary infrared dark cloud, $3.6\pm0.4$ kpc away in the Galactic plane containing a total of $\sim1000 M_{\sun}$ \citep{pe14}. 
It consists of four, parsec-long filaments, including SDC13.174-0.07, SDC13.158-0.073, and SDC13.194-0.073 \citep{pe09}, which converge onto a central hub. The IRAM 30m MAMBO 1.2 mm continuum data, with a resolution of 10.7\arcsec, identified 18 compact starless and protostellar cores distributed both along the filaments and within the central hub. Among these, the two most massive cores (named MM1 and MM2) are located at the junction of the four converging filaments within the central hub \citep{pe14}. \nh3 (1,1) and (2,2) line data from the Jansky Very Large Array (JVLA) and Green Bank Telescope (GBT), with a resolution of 4\arcsec, show significant radial and longitudinal velocity gradients along these filaments, with a velocity dispersion increasing toward the local density peaks and the hub center where the filaments spatially converge \citep{wi18}. In their work, this velocity structure is interpreted as the consequence of the gravitational collapse of gas along the filaments toward the center. Therefore, these filaments act as mass reservoirs to replenish the central hub, sustaining a density condition required for massive star/cluster formation.

In this paper, we report the continuum polarization observations toward SDC13, using the JCMT POL-2 polarimeter. These observations, with a physical resolution of $\sim$0.5 pc, allow us to probe the pc-scale magnetic field within SDC13 and investigate how the local magnetic field interacts with the density and kinematic structures in order to understand the current physical conditions of SDC13. By comparing with the large-scale magnetic field traced by $PLANCK$ polarization data, revealing the environment at the earlier evolutionary stage, we aim at studying the evolutionary history of the SDC13 HFS and determine its origin. In \autoref{sec:obs}, we present the observations and data reduction. \autoref{sec:results} reports the observed magnetic field with the JCMT POL-2. \autoref{sec:ana} presents how we estimate the filamentary density structures, the local velocity gradients, and the local gravitational force from the observed data. A statistical analysis is performed to identify possible trends and correlations between various physical parameters. 
With this, we study the local interplay and provide a global stability analysis. 
In \autoref{sec:dis}, we discuss how the observed smaller-scale features in the SDC13 HFS are connected to and have evolved from the large-scale environment.
Our conclusions are summarized in \autoref{sec:con}.



\section{Observations}\label{sec:obs}
We carried out polarization continuum observations toward SDC13 with the reference position (R.A., Dec.)=(18$^{h}$14$^{m}$28.5$^{s}$, -17\degr33\arcmin30\farcs9) with SCUBA-2 POL-2 mounted on the JCMT (project code M17BP041 and M19AP040; PI: Koch). Our target was observed in two epochs: 3 sets of 1-hour integration on August 17, 2017, and 28 sets of 30-minute integration on March 22, 2019, and May 19, 2019. All these observations were taken in band-1 weather with a $\tau_{225 GHz}$ opacity ranging from 0.03 to 0.04. The POL-2 DAISY scan mode \citep{fr16} was adopted, producing a fully sampled circular region with a diameter of 11\arcmin\ and a resolution of 14\arcsec. Both 450 $\mu$m and 850 $\mu$m continuum polarization were observed simultaneously. This paper focuses on the 850 $\mu$m data.

The POL-2 polarization data were reduced using $pol2map$\footnote{http://starlink.eao.hawaii.edu/docs/sc22.pdf} in the \textsc{smurf} package\footnote{version 2019 Nov 2} \citep{be05,ch13}. The reduction procedure followed the $pol2map$ script. The $skyloop$ mode was invoked in order to reduce the map-making uncertainty, and the MAPVARS mode was activated to estimate the total uncertainty from  the standard deviation among the individual observations. The details of the data reduction steps and procedure are described in a series of BISTRO papers \citep[e.g.,][]{wa17, kw18, wa19}. The POL-2 data reduction was done with a 4\arcsec\ pixel size, because larger pixel sizes can increase the uncertainty due to the map-making process. 

The output Stokes I, Q, and U images were calibrated in units of mJy/beam, using a flux conversion factor (FCF) of 725 mJy/pW \citep{de13}, and binned to a pixel size of 7\arcsec\ to improve the sensitivity and produce a Nyquist-sampled polarization map. We note that the atmospheric background removal technique used in the JCMT observations filters out extended source structures with scales $\gtrapprox$ 3\arcmin.

The uncertainty in POL-2 images originates from a combination of instrumental noise and uncertainty in the map-making. The DAISY scan mode generates a map with the lowest and nearly uniform instrumental noise within the central 3\arcmin-diameter region, and an increasing noise towards the edge of the map. The non-linear map-making process can enhance small perturbations in the input data and cause non-negligible differences in the resulting intensity maps. This is treated as an additional source of uncertainty. The typical rms noise of the final Stokes Q and U maps is $\sim$ 1.1 \mjyb\ at the center of the map. The Stokes I image has a higher intensity-dependent rms noise of 1--5 \mjyb\ near the central intensity peaks, as a result of larger map-making uncertainties. The calculated polarization fraction $P$ was debiased with the asymptotic estimator \citep{wa74} as
\begin{equation}\label{eq:debias}
P=\frac{1}{I}\sqrt{(Q^2+U^2)-\frac{1}{2}(\sigma_{Q}^2+\sigma_{U}^2)}
\end{equation}
with an uncertainty $\sigma_{P}$ calculated as
\begin{equation}\label{eq:dp}
\sigma_{P}=\sqrt{\frac{Q^2\sigma_{Q}^2+U^2\sigma_{U}^2}{(Q^2+U^2)I^2}+\frac{\sigma_{I}^2(Q^2+U^2)}{I^{4}}},
\end{equation}
where $\sigma_{I}$, $\sigma_{Q}$, and $\sigma_{U}$ are the uncertainties in the 
$I$, $Q$, and $U$ Stokes parameters.

\section{Results}\label{sec:results}
\begin{figure*}
\includegraphics[width=\textwidth]{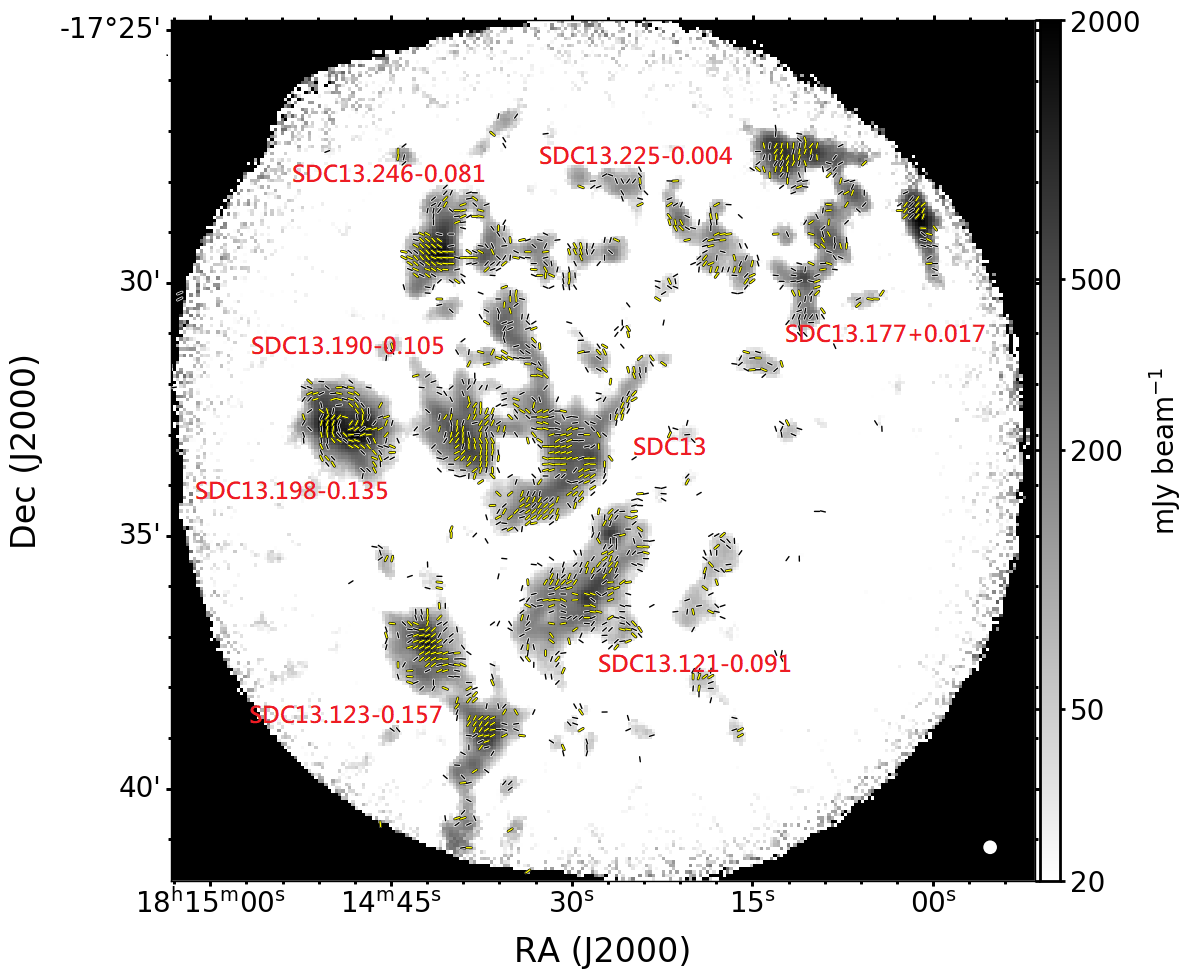}
\caption{B-field orientations (segments) sampled on a 7\arcsec\ grid overlaid on 850 $\mu$m dust continuum, sampled on a 4\arcsec\ grid, of the SDC13 region observed with POL-2. The yellow and black segments display the larger than 3$\sigma$ and 2--3$\sigma$ polarization detections, rotated by 90\degr\ to represent magnetic field orientations.  }\label{fig:Bmap_all}
\end{figure*}

\autoref{fig:Bmap_all} shows the observed 850 $\mu$m polarization map, overlaid on the Stokes $I$ image with a pixel size of 4\arcsec. Since the POL-2 observing mode filters out extended emission larger than a few arcminutes, the POL-2 continuum map only shows dense clouds with a physical scale of a few parsec. In the following these are referred to as $compact$ $clouds$, in order to distinguish them from the larger-scale diffuse clouds with angular sizes larger than a few arcminutes and column densities lower than $10^{22}$ cm$^{-2}$ that are further discussed in \autoref{sec:env10pc}. Our target SDC13 is the ``Y-shape" compact cloud located at the center of the field of view in \autoref{fig:Bmap_all}. In addition to SDC13, our observations also reveal the magnetic field morphologies within the nearby compact clouds. In this paper, our analysis focuses on the SDC13 system, for which high-resolution NH$_3$ line data are available to also trace the gas kinematics.

\subsection{Dust Continuum}
\begin{figure*}
\includegraphics[width=\textwidth]{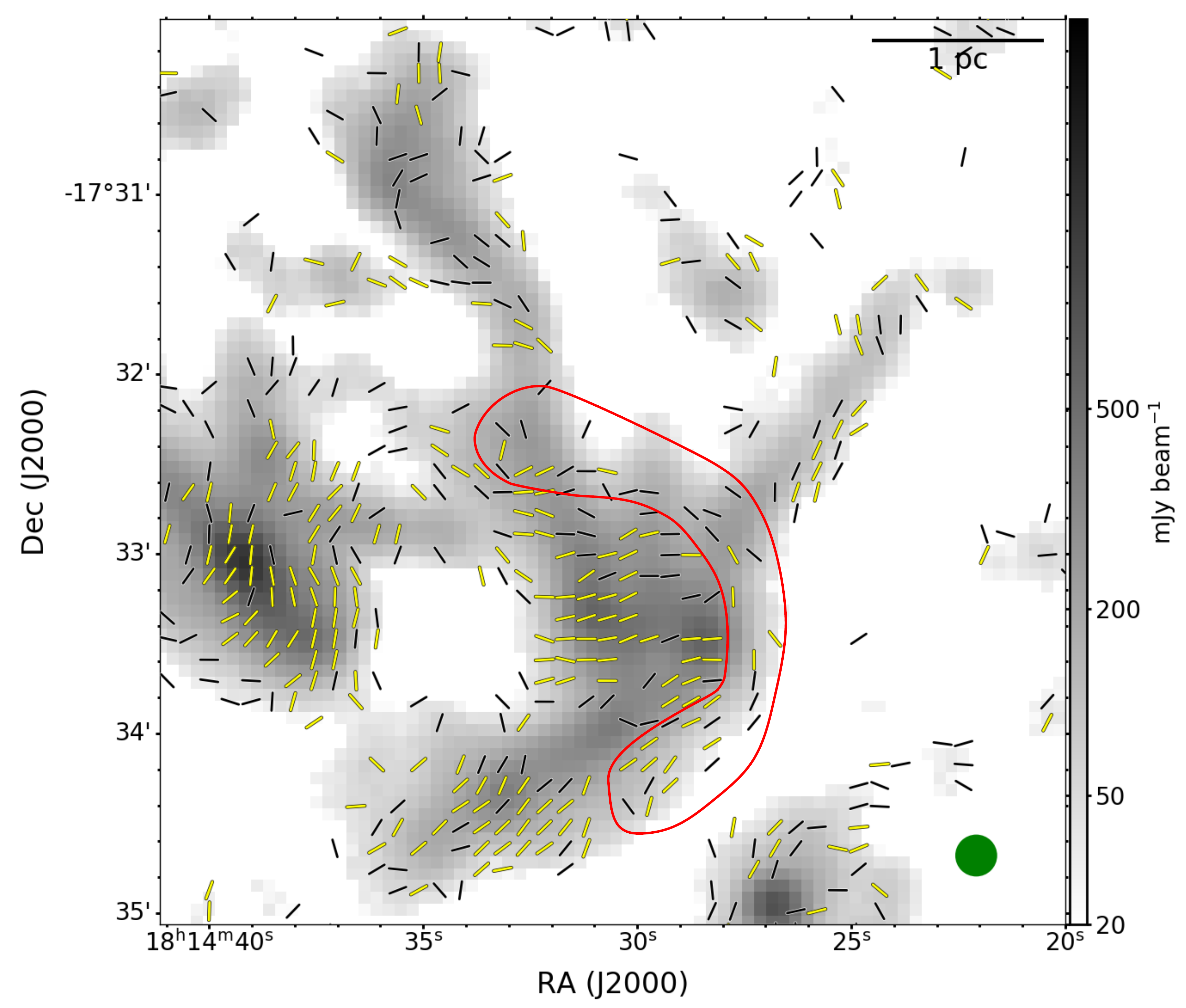}
\caption{Identical to  \autoref{fig:Bmap_all}, but zoomed-in on SDC13. The green circle at the right bottom is the JCMT POL-2 beam size. The red contour separates the arc-like "U-shape" magnetic field morphology which appears to be significantly 
influenced by the larger-scale surroundings (see later discussion sections).
}\label{fig:Bmap_SDC13}
\end{figure*}

\autoref{fig:Bmap_SDC13} presents a zoom-in polarization map toward SDC13.
The POL-2 850 $\mu$m continuum image shows a dense hub located at the center of SDC13, composed by one starless core and two protostellar cores. Four parsec-long filaments (north, north-west, north-east, and south) are converging to this central hub within SDC13, forming a typical hub-filament system. In addition to this major hub, a bright clump can be seen east of SDC13, possibly connected to SDC13 through two bridging filaments. We note that this eastern clump is bright in 8 $\mu$m $Spitzer$ GLIMPSE \citep{ch09} and 24 $\mu$m $Spitzer$ MIPSGAL \citep{ca09} maps, which might suggest that the eastern clump is possibly more evolved than SDC13.

\subsection{Polarization Data Selection}
In order to ensure significant detections, we select polarization segments with the criteria $I/\sigma_{I}>10$ and $P/\sigma_{P} >2$. In this way, a total of 1130 polarization measurements are selected over the field of view, including 573 segments with $P/\sigma_{P} >3$ and 557 segments with $3>P/\sigma_{P} >2$. These segments are not only distributed within SDC13, but a significant numbers 
are in the nearby compact clouds. 
Our analysis in the following section will focus on the polarization segments in SDC13, and we will further investigate the polarization patterns in the nearby compact clouds when discussing the larger-scale surroundings of SDC13 in \autoref{sec:env10pc}.
In an attempt to increase the statistics for the analyses, we note that 
\autoref{fig:Bmap_all} shows that the orientations of the $3> P/\sigma_{P}>2$ and $P/\sigma_{P} >3$ polarization segments are locally similar. This suggests that the $3> P/\sigma_{P}>2$ segments can be used to fill in gaps. Hence, we will also adopt 
these data in our analyses in order to be able to work with more connected B-field structures.

For the final selected data, $P/\sigma_{P} >3$ data have a maximum $\sigma_{PA}$ of 12.0\degr\ with a median $\sigma_{PA}$ of 7.2\degr. $3>P/\sigma_{P} >2$ data have a maximum $\sigma_{PA}$ of 15.3\degr\ with a median $\sigma_{PA}$ of 10.7\degr. 
\autoref{sec:pp} discusses
whether the observed polarization traces magnetically aligned dust grains and how the polarization fraction correlates with the total intensities, thus tracing the column densities.
Based on a Bayesian analysis it is found that, indeed, dust grains are likely magnetically aligned in SDC13.

\subsection{Magnetic Field Morphology}
\autoref{fig:Bmap_SDC13} presents the POL-2 850 $\mu$m continuum Stokes $I$ image, with a pixel size of 4\arcsec\, toward SDC13 with the Nyquist-sampled (7\arcsec\ ) magnetic field detections, rotated by 90\degr\ with respect to the detected polarization orientations assuming that the observed polarization traces magnetically aligned dust grains \citep[e.g.,][]{cu82,hi88}.
The overall magnetic field morphology is spatially variant. In the central hub, the magnetic field is almost uniform with an east-west orientation (PA $\sim$ 90\degr). The magnetic field is turning northwest-southeast (PA $\sim$ 100-140\degr) as it approaches the northwestern and the southern filament, and it remains like this over extended parts along these filaments. On the other side of the hub, the field is
hinting a northeast-southwest orientation (PA $\sim$ 30-50\degr) in some sections along the northeastern filament. 

In addition to the overall smoothly varying magnetic field morphology, a narrow arc-like, 
"U-shape"\footnote{In the following sections, the terms "arc-like" and "U-shape" are used synonymously.
} magnetic field structure is depicted at the western edge of the central hub (marked as a red area in \autoref{fig:Bmap_SDC13}). Here, B-field orientations change rather abruptly from the nearby regions. 
This pattern is commonly seen in numerical simulations of magnetic fields compressed by shocks or gas flows \citep[e.g.,][]{in13,go18}. This arc-like structure appears to be connecting regions with no polarization detections or polarization holes. Hence, these polarization holes might be caused by depolarization, 
resulting from underlying more complicated B-field structures caused by compression, shocks, and turbulence in gas flows.

\begin{figure*}
\includegraphics[width=\textwidth]{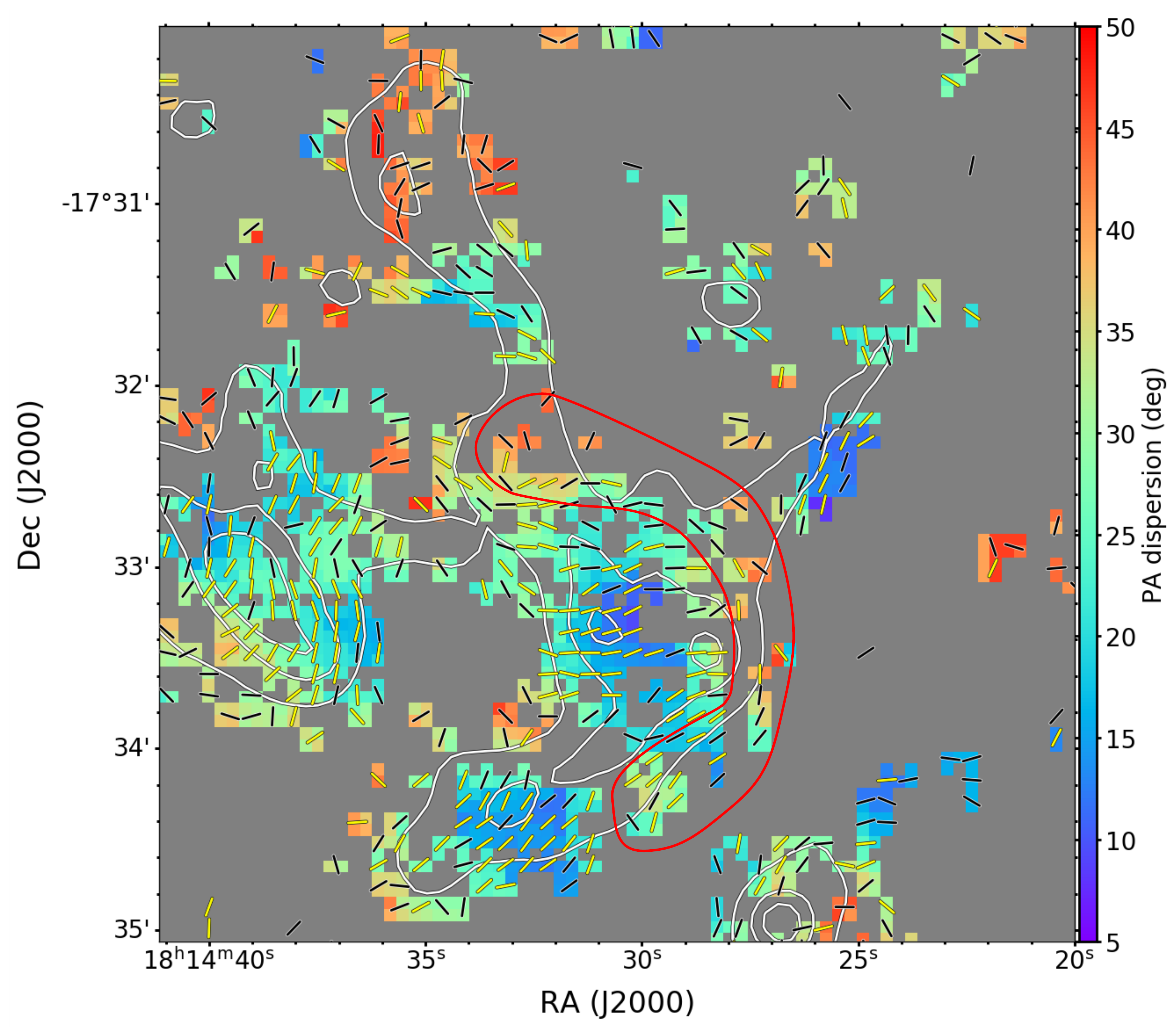}
\caption{Local magnetic field angular dispersion map shown in color scale. The angular dispersion is calculated using segments within an 18 \arcsec radius for each pixel.
Pixels with less than 3 nearby segments are excluded. The white contours are the 100, 300, and 500 \mjyb intensity of the 850 $\mu$m continuum. The yellow and black segments display the magnetic field orientations as in \autoref{fig:Bmap_SDC13}. The red contour marks the arc-like "U-shape" magnetic field morphology.
The magnetic field is relatively uniform within the central hub and individual filaments, leading to small dispersion values. It becomes more disordered near the arc region and around the dense core in the northeastern filament, leading to larger local dispersion values.
}\label{fig:BDISmap_SDC13}
\end{figure*}

In order to investigate how the magnetic field morphology in this arc-like structure delineates itself from the central hub and the connecting filaments, we show a magnetic field angular dispersion map in \autoref{fig:BDISmap_SDC13}.
This local dispersion is calculated using polarization segments within a radius of 18\arcsec (1.5 beam size), for each pixel,
and it hence is a measure for how much a field orientation changes from its near surroundings (see, e.g., \citet{ko18,fi16,plxx,plxix}).
The resulting local dispersion appears to decrease toward the dense cores and the central hub from $\sim25\degr$ to $\sim15\degr$. Different from that, the arc-like structure has patches of larger local dispersion values ($>30\degr$), separating it from its near surrounding.
Since this arc-structure spans a few parsec with distinctively larger dispersion values, it is probably a separate feature and possibly originated from cloud-scale kinematics events. 
We, nevertheless, acknowledge that the incomplete polarization coverage
prevents us from identifying this as a fully connected and coherent structure, and its endpoints are not uniquely defined.
This feature is further discussed in the context of the large-scale environment in \autoref{sec:env10pc}.

\section{Analysis}\label{sec:ana}
In this section we analyze the physical properties of SDC13 from both $local$ and $global$ perspectives, following the analysis scheme in \citet{wa21}. \autoref{sec:ana_local} aims at extracting the spatial properties within SDC13, including the filamentary structures, magnetic field, gravitational field, and gas velocity gradient. We investigate how these physical quantities possibly correlate with each other {\it locally}, in order to discuss the possible mechanisms that lead to the formation and evolution of SDC13. In contrast to that, \autoref{sec:ana_global} focuses on investigating the global stability of hub and filaments in SDC13, revealing whether these systems are {\it globally} stable or collapsing.

\subsection{Local Interplay among Filaments, Magnetic Field, Gravity, and Gas Kinematics}\label{sec:ana_local}
\subsubsection{Filament Identification}\label{sec:filament}
\begin{figure*}
\includegraphics[width=\textwidth]{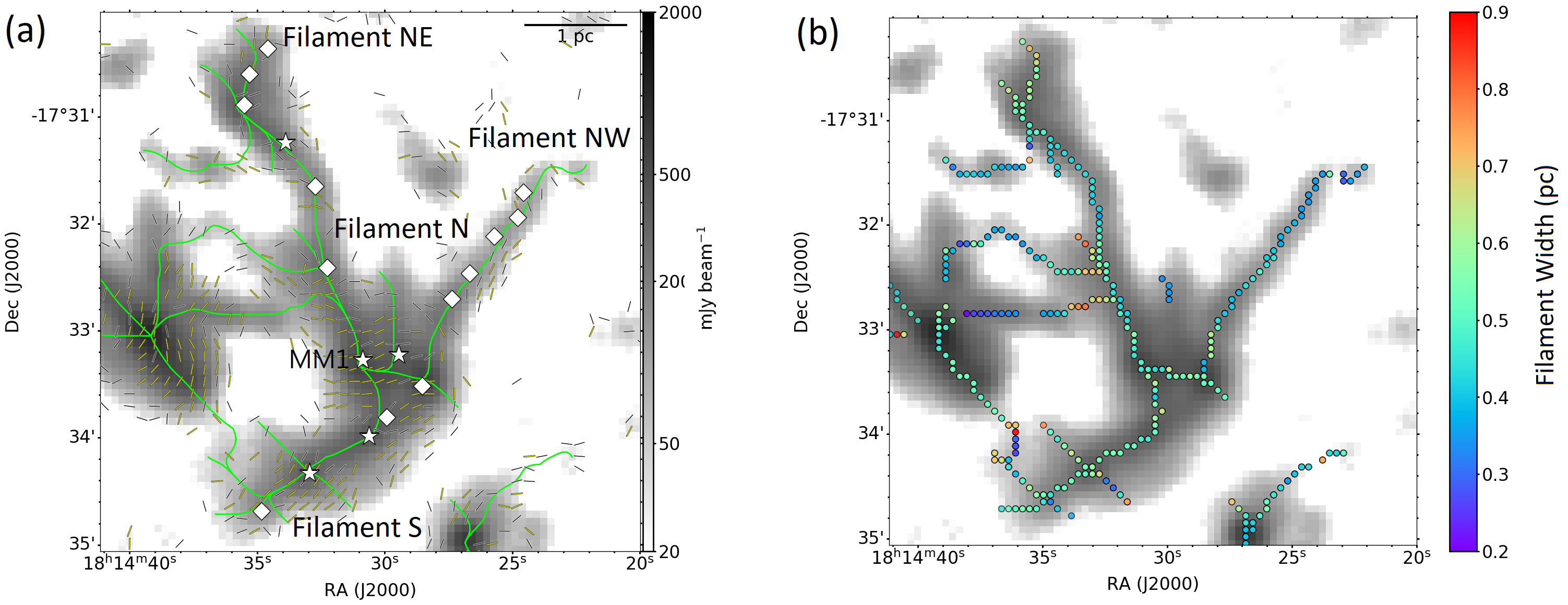}
\caption{Identified filaments overlaid on the 850 $\mu$m dust continuum map. (a) The green lines are the filaments longer than 1\arcmin.
The white diamonds and stars label the starless and protostellar sources identified in \citet{pe14}. The yellow and black segments display the larger than 3$\sigma$ and 2--3$\sigma$ polarization detections, rotated by 90\degr\ to represent magnetic field orientations. (b) Widths of the identified filaments. The filament widths increase with local density, from $\sim$0.25 pc, close to the JCMT beam size, in the diffuse regions to $\sim$0.5 pc within the dense hub.}\label{fig:Fmap}
\end{figure*}

In order to identify the ridges of filamentary structures in the JCMT 850 $\mu$m Stokes I image, we adopt the $DisPerSE$ algorithm \citep{so11}. We use a contrast threshold of 20 \mjyb ($\sim$5$\sigma$) for the filament identification, and exclude the identified filaments with lengths shorter than 1 arcmin to ensure the significance. The identified filaments are plotted in \autoref{fig:Fmap}(a). The four longest filaments, northeast (NE), northwest (NW), north (N), and south (S) filament are shown to converge to the central hub, forming a ``Y''-shaped hub-filament system. This is consistent with the four filaments identified in \citet{pe14} from IRAM 30m MAMBO 1.2 mm dust continuum data and also with the \nh3 observations in \citet{wi18}. Several shorter filaments are merging into the longer filaments via local convergent points. Two bridging filaments seem to connect the filament NE and the eastern hub.

We use the python package $FilChap$ \citep{su19} to estimate the widths of the identified filaments. We fit individual radial intensity profiles extracted at each pixel position along the filament ridges. A bootstrap method is applied to fit these radial profiles with a Gaussian function, and the uncertainties are estimated using a Monte Carlo approach to generate 100 simulated profiles based on the observed intensities and uncertainties. Those fits with a width smaller than three times the uncertainties are excluded from the further analysis. The fitted Gaussian widths ($\sigma_F$) are converted to the FWHM width ($\Delta F$) via $\Delta F= \sqrt{8ln2} \sigma_F$ which are shown in \autoref{fig:Fmap}(b). We note that the intensity profiles along the filament N overlap with filament NE and NW leading to highly uncertain fitting results.
They are, therefore, not shown in \autoref{fig:Fmap}(b).
Generally, the filament widths grow with the local intensity, from $\sim$0.3 pc in the diffuse regions to $\sim0.5$ pc near the dense hub and clumps. We note that the smaller filament widths in the diffuse regions might be consistent with a universal 0.1 pc filament width discovered by the $Herschel$ Gould Belt survey \citep[e.g.,][]{ar11}, although our resolution of 0.24 pc is insufficient to clearly resolve a 0.1 pc width.

In order to investigate the occurrence of star formation in the SDC13 hub-filament system, we overplot the starless and protostellar cores from \citet{pe14} on the filaments in \autoref{fig:Fmap}. Without any exception, all the cores are located either on filament spines or in filament convergent points. The protostellar cores are mostly found near convergent points, while more starless cores are distributed along the filaments. T  his distribution is suggestive of star formation taking place not only in the central hub, but also along the converging filaments. Moreover, local convergent points seem to have a higher probability of triggering star formation. These features have also been seen in \citet{wi18} where the filaments are identified from the \nh3 data. This suggests that \nh3 and continuum data likely trace the same filamentary structures.

\subsubsection{Local Gravitational Field}\label{sec:gravity}
In order to investigate whether gravity influences the formation of the converging filaments, we estimate the projected gravitational vector field from the JCMT 850 $\mu$m continuum data. Since the large-scale emission
beyond about 3\arcmin
is filtered out by the JCMT POL-2 observations, we focus here on the local gravitational field within the denser SDC13. 
The inter-cloud scale gravitational field, traced by $Herschel$ observations, is discussed In \autoref{sec:env10pc}. 

Following the development of the polarization-intensity gradient technique in \citet{koch12a,koch12b}, 
the local projected gravitational force acting at a pixel position ($\vec{F_{G,i}}$) can be expressed as the vector sum of all gravitational pulls generated from all pixel positions over the map \citep{wa21} as
\begin{equation}
\vec{F_{G,i}} = kI_i\sum_{j=1}^{n}\frac{I_j}{r_{i,j}^2}\hat{r},
\end{equation}
where $k$ is a factor accounting for the gravitational constant and conversion from emission to total column density. $I_i$ and $I_j$ are the intensity at the pixel position $i$ and $j$, and $n$ is the total number of pixels within the area of relevant gravitational influence. $r_{i,j}$ is the plane-of-sky projected distance between the pixel $i$ and $j$, and $\hat{r}$ is the corresponding unity vector. 
The above equation assumes that the intensity distribution is a fair approximation for the distribution of the total mass, and that the mass components in SDC13 are roughly at the same distances.
A constant factor $k=1$ 
is used because we will only utilize the directions of the local gravitational forces and not their absolute magnitudes. When calculating the local gravitational field,
a threshold of 20 \mjyb ($\sim5\sigma$) is introduced below which any gravitational influence is neglected. This is justified because any gravitational force originating from diffuse and extended structures tends to be rather symmetrical which means that any already small gravitational pulls will largely cancel out. 

\begin{figure}
\includegraphics[width=\columnwidth]{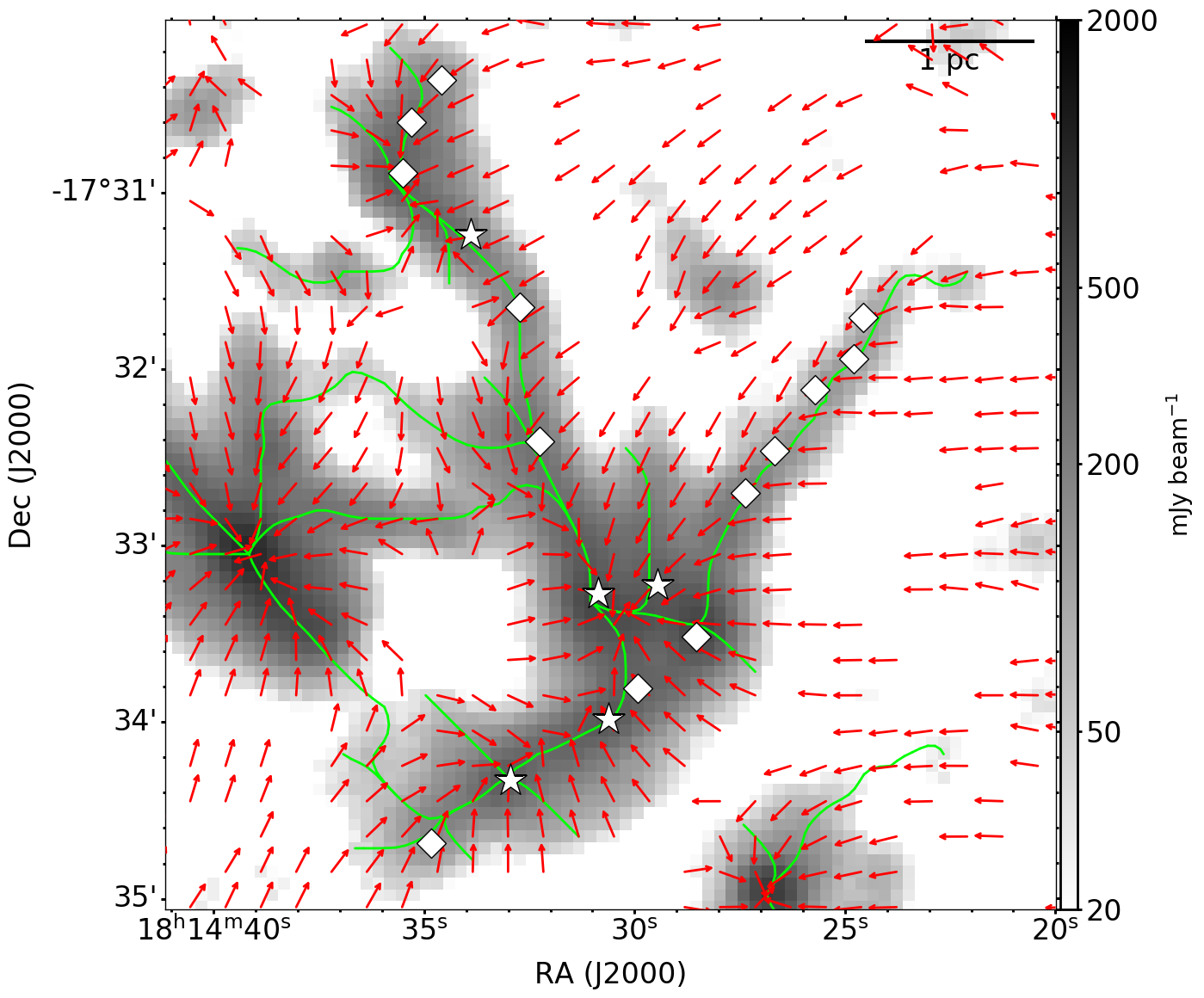}
\caption{Projected local gravitational vector field overlaid on the 850 $\mu$m dust continuum map. The red arrows represent the direction of the local gravitational field, only plotted per independent beam. The green lines are the filaments longer than 1\arcmin. The white diamonds and stars label the starless and protostellar sources identified in \citet{pe14}.}\label{fig:Gmap}
\end{figure}

\autoref{fig:Gmap} displays the local gravitational vector field in SDC13. Looking for distinct directions in this vector field,
local gravity visually appears as a combination of two modes: (1) pulling to the filament ridge (local vectors prevailingly orthogonal to the filament's ridge) and (2) pulling to a converging center (local vectors rather along the filament's ridge and local vectors pointing azimuthally symmetrically to a mass center. 

The relative importance of these two modes likely determines the orientation offsets between local gravity and filaments. In the NE and S filament, local gravity seems to be more efficiently dragging material orthogonally to the filament ridges, except for the two local converging centers at the tips of the filaments. 
In the NW filament, the gravitational force pulling toward the central hub is likely comparable to the force toward the filament ridge, and hence an orientation offset of $\sim$30--60\degr can be seen between local gravity and filament. In the hub center, the gravitational field is mainly pointing toward the most massive protostellar core within the central hub (also known as MM1), and the orientation offsets between local gravity and filaments are $\sim$40--90\degr. We note that this analysis is limited by the observational resolution, and
the gravitational field originated from possible structures at scales smaller than our resolution cannot be directly probed.

\subsubsection{Local Velocity Gradient}
\begin{figure}
\includegraphics[width=\columnwidth]{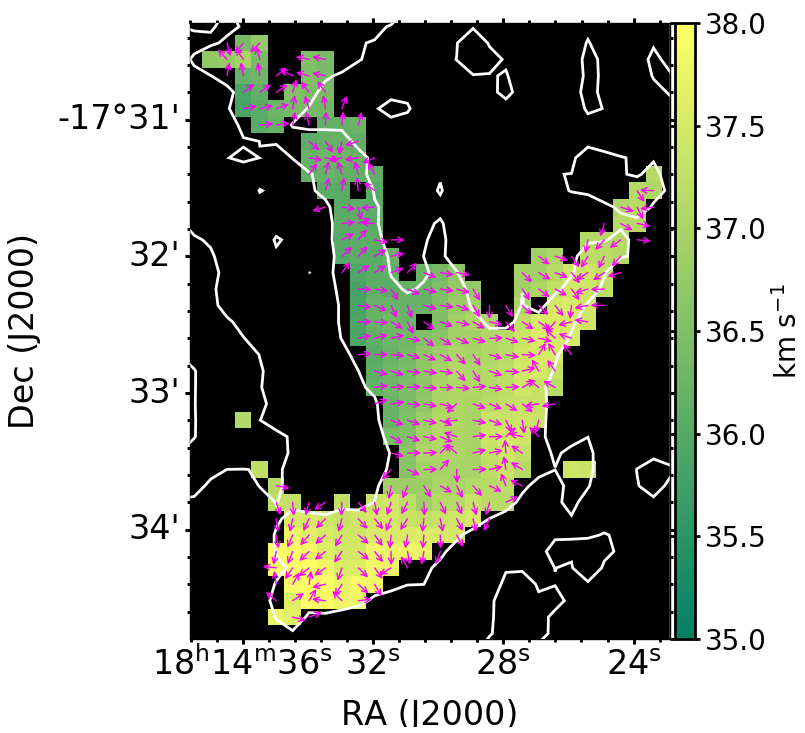}
\caption{\nh3 local velocity gradient field overlaid on the centroid velocity map. The local velocity gradients are calculated using the centroid velocity map smoothed to a pixel size of 7\arcsec, 
which is comparable to the pixel size of our POL-2 data. The white contour represents the \nh3 integrated intensity of 0.2 Jy/beam~km/s. The magenta arrows show the local velocity gradients with constant lengths to emphasize their directions. Near the central hub, the velocity gradients mostly show an east-west direction, while the local velocity gradients in the filament regions are either converging to the local clumps or to the filament ridge. }\label{fig:vgmap}
\end{figure}

To analyze the velocity structure in SDC13, we derive local velocity gradients from the \nh3 (1,1) centroid LOS velocity map in \citet{wi18}. Probing the velocity structures at the same physical scale as the magnetic field traced by our polarization data requires smoothing the \nh3 data to a pixel size of 7\arcsec\ from the original beam size of 4.0\arcsec$\times$2.8\arcsec (\autoref{fig:vgmap}). We estimate the centroid velocity by using the fitting schemes in the $CLASS$ software\footnote{The fitting procedure for hyperfine structures is described at: https://www.iram.fr/IRAMFR/GILDAS/doc/pdf/class.pdf}. The {\it local} velocity gradient is calculated by fitting the 2D spatial distribution of the centroid velocities within 3$\times$3 pixel \citep{go93} with
\begin{equation}\label{eq:vg}
V_{LSR}= c_{x}x + c_{y}y + c_0,
\end{equation}
where x and y are the positions of each pixel, and $c_0$, $c_{x}$, and $c_{y}$ are free parameters representing the first-order expansion of the velocity field. 

The resulting local velocity gradient field is plotted in \autoref{fig:vgmap}.
The velocity gradients are along an east-west direction near the hub center. They become more complex in the filament regions, seemingly either pointing toward the local clumps along the filaments, or being perpendicular to the filament ridges. A more thorough statistical analysis is necessary to understand whether these trends are significant or not.

\subsubsection{
Possible Correlations and Trends among Filaments, Magnetic Field, Gravity, and Gas Kinematics}\label{sec:correlation_region}
In this section
we perform systematic statistical analyses to
investigate how the physical parameters act in the hub-filament system (HFS) and to 
reveal possible correlations among the orientations/directions of filaments (F), magnetic field (B), gravitational force (G), and gas velocity gradient (VG). The magnitudes of these parameters are excluded from our further analysis here, because our focus is on understanding first the possible correlations in orientations.
Moreover, the complete 3-dimensional information to possibly correct the projected magnitudes is observationally not accessible.

For the selected filaments within SDC13, we estimate a filament orientation at every pixel in the following way.
For the $ith$ pixel along the ridge of a filament, we fit the positions of the $(i-2)th$ to the $(i+2)th$ consecutive pixels along the filament with a straight line to estimate the local filament orientation. In this way, the median fitting error in the local orientation of a filament is $\sim$2\degr.

\begin{figure*}
\includegraphics[width=\textwidth]{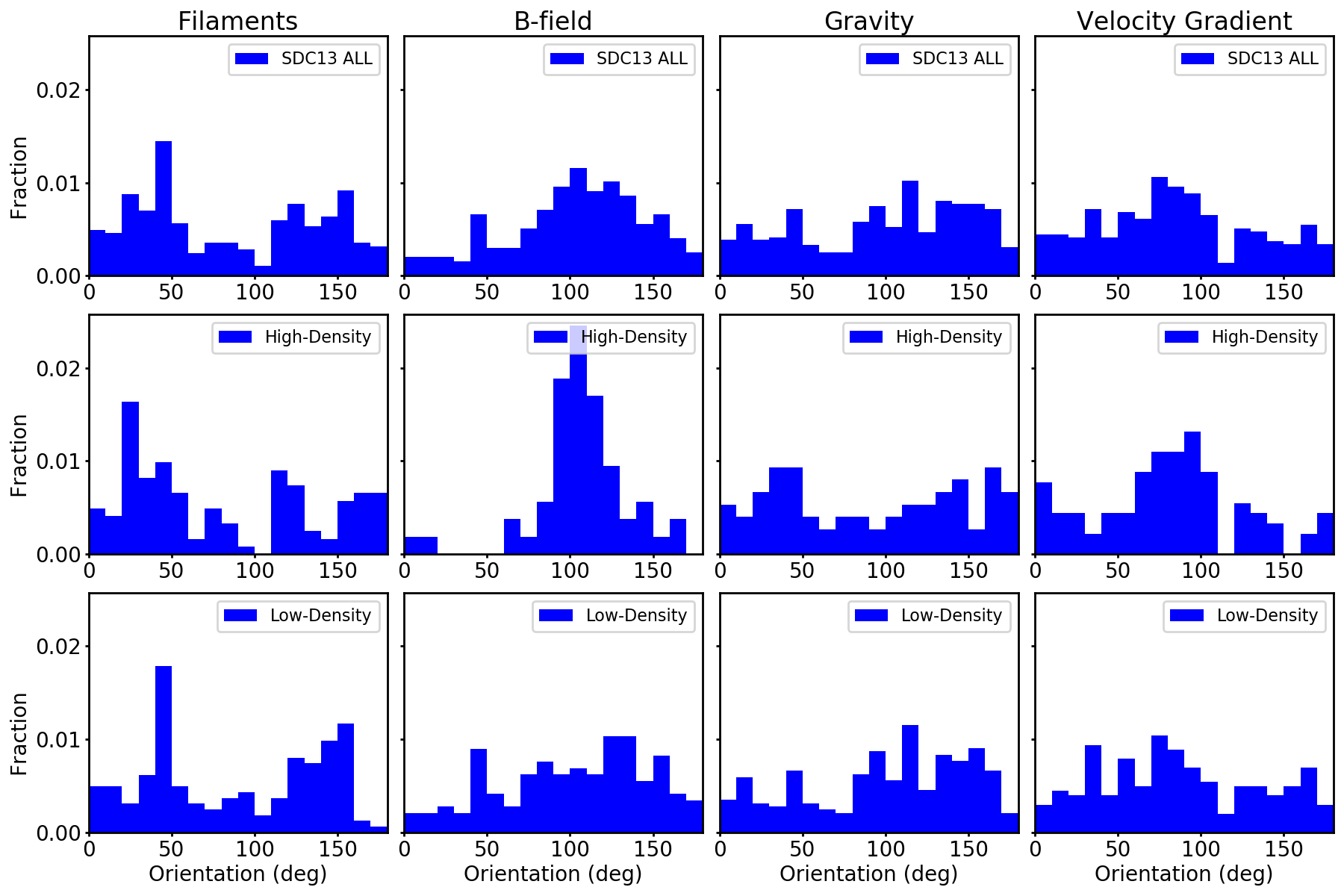}
\caption{Histograms of filament, magnetic field, local gravity, and local velocity gradient directions/orientations (columns) for the entire SDC13 (top), high-density ($>$250 mJy/beam; middle), and low-density regions ($<$250 mJy/beam; bottom). The fraction shown in the y-axis is the probability density, and thus the integral of each histogram is normalized to unity.
}\label{fig:hist_all}
\end{figure*}

\autoref{fig:hist_all}
presents the histograms of the local orientations of filaments (F), magnetic field (B), gravity (G), and velocity gradients (VG) for the entire SDC13, and separated into high- and low-density regions.
In all the regions, F, B, G, and VG show rather clear single or multiple peaks in their distributions. This is clearly different from random, i.e., uniform distribution in orientations.
Noticeably, the magnetic field orientations display a more pronounced peak in high-density regions. Other parameters also appear to change with density, though in less
definite ways. 

To further isolate possible correlations, we perform 
an all-pairwise comparison of the four parameters (F, B, G, VG), looking at their relative orientations at different densities.
For each measurement of one parameter, we select a nearest measurement of another parameter within a radius of 14\arcsec (one independent beam). These two measurements are then defined as one associated pair for the two parameters. A relative orientation ($\Delta PA$) is calculated for each associated pair. We perform a one-sample Kolmogorov–Smirnov (KS) test to the resulting six $\Delta PA$ distributions in each region, to examine whether these distributions differ from a uniform distribution, as the null hypothesis. For each distribution, a probability (p-value) that the observed distribution can be drawn from the hypothetical uniform distribution is calculated, and a threshold of p=0.05 (95\% confidence interval) is used to reject the null hypothesis. We note that a p-value higher than 0.05 does not automatically indicate a random distribution, but could also result from an insufficient number of data points. \autoref{tab:pair} lists the features of those parameter pairs with non-random distributions. The complete histograms for all the parameter pairs, additionally divided into hub and filament regions, are given in \autoref{sec:allpair}.

\begin{figure*}
\includegraphics[width=\textwidth]{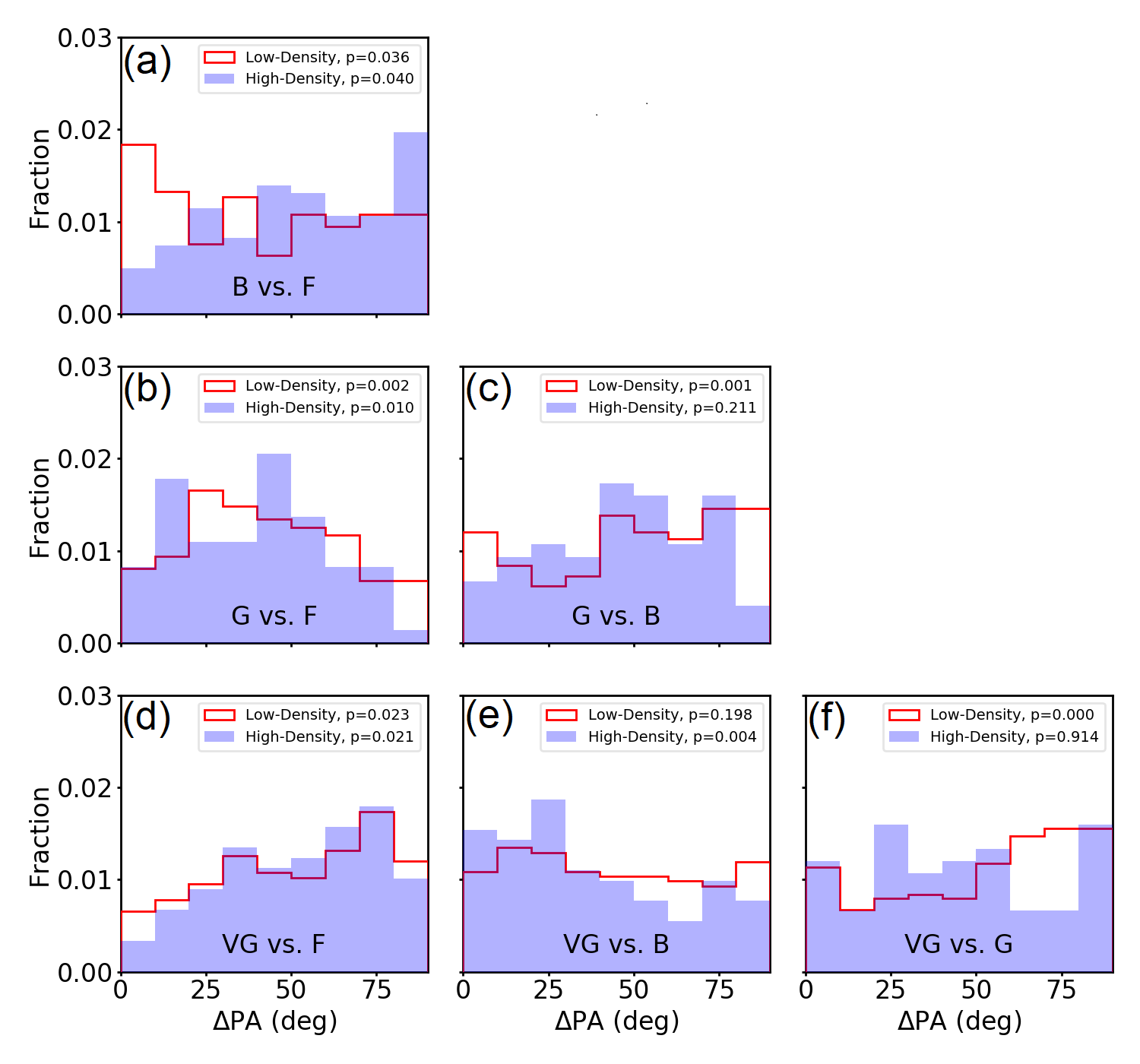}
\caption{Histograms of pairwise relative orientations among filaments, magnetic field, local gravity, and local velocity gradients, 
separated into diffuse and dense regions with an intensity threshold of 250 mJy/beam.}\label{fig:hist_2I}
\end{figure*}

Since the dominating physical process within a cloud might evolve with local density \citep{so13,pl16,wa21}, 
we have introduced a density threshold
of $2\times10^{22}$ cm$^{-2}$ (250 mJy/beam, assuming a dust temperature of 27.2 K \citep{wi18}),
defining low- and high-density regions. This threshold is motivated as the differences in the resulting $\Delta PA$ distributions become most obvious. The histograms of the relative orientations for these two regimes are shown in \autoref{fig:hist_2I}.
The most significant difference between these two regimes is that filaments tend to be more aligned with the magnetic field in low-density regions, but become more perpendicular in high-density regions (\autoref{fig:hist_2I}(a)). Similar trends are seen in numerous molecular clouds based on \textit{PLANCK} data \citep{pl16},
though with a substantially larger beam these data are rather probing the transition between the diffuse interstellar medium and molecular clouds.

In addition to the magnetic field trend, both  gravity and velocity gradients appear to
fall into a similar range of relative orientations, $\sim$20--75\degr, 
with respect to the filaments (\autoref{fig:hist_2I}(b) and (d)). 
This offset angle is likely representing the combination of the two modes in the local gravitational force map, as pointed out in 
\autoref{sec:gravity}.
Gravity can locally both pull the gas toward a filament ridge (leading to a more perpendicular gravity-filament configuration) and also radially toward the central hub (leading to a more parallel gravity-filament configuration). The offset angle is likely determined by the relative importance of these two modes. Thus, it can still vary over a range. Similarly, the offset angles between the velocity gradients and the filaments might be linked to these two modes. We, however, note that the G--F offset angle is not necessarily the same as the VG--F offset angle, because a filament's radial and longitudinal collapsing timescale can be different, depending on the density and the geometry of a filament \citep{va09,va17}.

An additional finding is that the velocity gradient in the low-density regions is more 
correlated with gravity (being more perpendicular), while in the high-density regions it is more correlated with the magnetic field (being more parallel; \autoref{fig:hist_2I}(e) and (f)).
This growing alignment between magnetic field and velocity gradient might point at a role of the magnetic field in guiding gas motions
as density increases.
This possibly results from the enhanced magnetic field in the central hub, where the magnetic field shows the largest field strength and a small angular dispersion (\autoref{fig:BDISmap_SDC13}; \autoref{tab:CF}). 
In contrast to that, the magnetic field in the low-density regions, where overall the angular dispersion is larger, is more likely to be locally distorted by turbulence, core fragmentation, or external pressure. 
As a consequence, this more
complex and less organized magnetic field morphology presents a largely random orientation with respect to the velocity gradient. Therefore, the magnetic field is also less capable of constraining gas motion. 

Finally, we acknowledge that some of the possibly emerging correlations and trends discussed here can be affected by unknown projection and integration effects. A more complete statistical analysis with a larger sample is needed to fully establish these trends. 


\begin{deluxetable}{ccc}
\tablecaption{
Trends and KS-test Results of Physical Parameters for Two-density Regimes in SDC13.\label{tab:pair}}
\renewcommand{\thetable}{\arabic{table}}
\tablenum{1}
\tablehead{\colhead{Pairs} & \colhead{Low-density Regions} & \colhead{High-density Regions}}
\startdata
\hline
B vs. F & $\parallel$(0.04) & $\perp$(0.04)  \\
G vs. F &  $0-60\degr(0.002)$ & $20-60\degr(0.01)$ \\
G vs. B & $\perp$(0.001) & ... \\
VG vs. F & $50-75\degr(0.02)$ & $50-75\degr(0.02)$  \\
VG vs. B &  ... & $\parallel(0.004)$  \\
VG vs. G & $\perp(<0.001)$ & ...  \\
\enddata
\tablecomments{
p-values from KS-tests (in parentheses) for all pairs where p$<$0.05, i.e., a larger-than 95\% probability for a distribution to be different from random. 
The related histograms are shown in \autoref{fig:hist_2I}.
Possible ranges and trends for relative orientations are noted ($\perp$: perpendicular; $\parallel$: parallel).
}
{\addtocounter{table}{-1}}
\end{deluxetable}

\subsection{Global Stability}\label{sec:ana_global}
In this section, we aim to investigate the global stability of the filaments and the hub region in SDC13 by evaluating the balance between gravitational, magnetic, and kinetic energy. In order to estimate the magnetic field strength from the polarization data, we use both the Davis-Chandrasekhar-Fermi (DCF) method \citep{da51,ch53} in \autoref{sec:DCF}, and the Skalidis \& Tassis (ST) method \citep{sk21} in \autoref{sec:ST}. The calculated magnetic energy is compared with the gravitational and kinetic energy using the virial theorem in \autoref{sec:Virial}.

\begin{figure}
\includegraphics[width=\columnwidth]{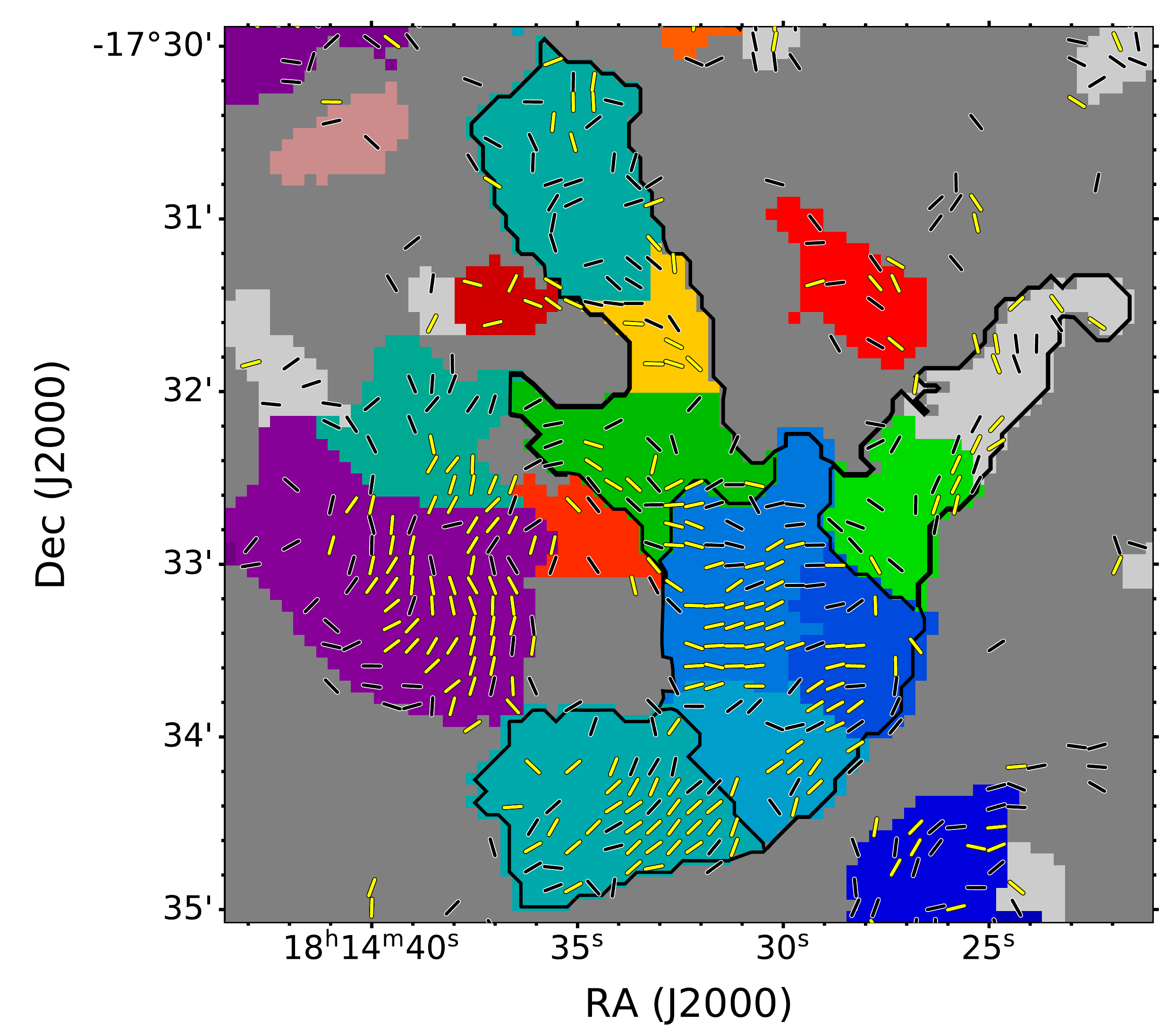}
\caption{Clumps identified in SDC13 using the ClumpFind algorithm. We group these clumps as central hub region (3 clumps), filament south (1 clump), filament north-east (3 clumps), and filament north-west (2 clumps) region, as labeled by the black lines, for the further local analysis. }\label{fig:regions}
\end{figure}

Since the stability of the hub and the filaments is possibly different, we separate SDC13 into hub and filament regions for the statistical analyses. 
We first use the $FindClump$ task in the $CUPID$ package \citep{be07} with the ClumpFind algorithm \citep{wi94} to separate SDC13 into individual clumps and then define the boundaries of each region (\autoref{fig:regions}). We group the three clumps in the center together as the hub regions, because they contain a group of dense cores and share similar polarization patterns. The three, two, and one clump(s) in the northeastern, northwestern, and southeastern side of the hub, respectively, are grouped together into the filament NE, WE, and S regions. The remaining clumps are excluded from further analyses here because they lack associated velocity data (\autoref{fig:vgmap}).

\subsubsection{Davis-Chandrasekhar-Fermi Method}\label{sec:DCF}
The DCF method assumes that kinetic and magnetic energy are in equipartition. Hence, the level of magnetic field perturbation, traced by the polarization angular dispersion $\delta \phi$, is the result of transverse incompressible turbulent Alfv\'{e}n waves, traced by the LOS non-thermal velocity dispersion $\sigma_{v,NT}$. Under this assumption, the plane-of-sky magnetic field strength ($B_{pos}$) can be estimated as
\begin{equation}\label{eq:CF}
B_{pos}=Q~\sqrt[]{4\pi \rho}\frac{\sigma_{v,NT}}{\delta \phi},
\end{equation}
where $\rho$ is the gas volume density, and Q is a factor accounting for complex magnetic field and inhomogeneous density structures. \citet{os01} suggested that $Q=0.5$ yields a good estimation of the magnetic field strength in the plane of sky if the magnetic field angular dispersion is less than 25\degr .

Since the role of the magnetic field is likely different in the filament and the hub regions of SDC13, we estimate the magnetic field strength in each region separately. 
Polarization segments are selected for each region, and a polarization angular dispersion is calculated.
In order to obtain a polarization dispersion originating from turbulent perturbation solely,
i.e., without any confusion from larger-scale magnetic field features, we calculate differences between polarization position angles using only nearest pixel pairs. In this way, the polarization angular dispersion is derived as
\begin{equation}
\delta \phi = \sqrt{\frac{1}{N-1}\sum_{i=1}^{N}\delta PA_{i}^2},
\end{equation}
where $\delta PA$ is the absolute difference between two polarization position angles for every possible nearest-polarization pair. N is the total number of pairs, and the factor $1/(N-1)$ is to debias the population standard deviation estimator. The calculated $\delta \phi$ for each region is listed in \autoref{tab:CF}.

To estimate the mean volume density in each region, we first construct a column density map using the POL-2 850 $\mu$m continuum data assuming a constant temperature of 12.7 K, adopted from the mean \nh3 rotational temperature \citep{wi18}, and a dust opacity $\kappa$ of 0.012 cm$^2$/g at 850 $\mu$m \citep{hi83}. The total mass of a region is then obtained by integrating the column density over the selected region. The volume of the hub region is estimated assuming a sphere with a FWHM diameter of 1.0$\pm$0.1 arcmin, obtained from a 2D-Gaussian fit to the hub. The volume of a filament region is derived adopting a cylinder with the filament's radius of 0.5 pc, given in \autoref{sec:filament}. The mean volume densities are then calculated from the total masses and their respective volumes.

In order to obtain a mean non-thermal velocity dispersion, we first average the observed \nh3 line widths in each region. Assuming a gas kinematic temperature ($T_\mathrm{kin}$) of 12.7 K, 
the thermal velocity dispersion for \nh3 is $\sqrt{\frac{k_B T_\mathrm{kin}}{m_{\mathrm{NH}_{3}}}}=0.10$ km~s$^{-1}$. The thermal velocity dispersion is then removed from the observed line width to obtain the non-thermal velocity dispersion ($\sigma_{v,NT}$) as
\begin{equation}\label{eq:vdisp}
\sigma^2_{v,NT}=\sigma^2_{obs} - \frac{k_B T_\mathrm{kin}}{m_{\mathrm{NH}_{3}}}
\end{equation}
where $\sigma_{obs}$ is the observed \nh3 velocity dispersion, and $m_{\mathrm{NH}_{3}}$ is the molecular weight.
The resulting magnetic field strengths for the four regions in SDC13 are listed in \autoref{tab:CF}.

The mass-to-flux criticality ($\lambda_{obs}$) is commonly used to evaluate the relative importance between magnetic field and gravity \citep{na78}, and calculated as
\begin{equation}
\lambda_{obs}=2\pi \sqrt{G}\frac{\mu m_{H}N_{H_2}}{B_{pos}},
\end{equation}
where $\mu$=2.33 is the mean molecular weight per H$_2$ molecule, 
$G$ is the gravitational constant, and $N_{H_2}$ is the molecular hydrogen column density. To correct for the unknown projection effect, \citet{cr04} suggest that a statistical average factor of $1/3$ can be used to better estimate the mass-to-flux ratio of oblate spheroid cores, flattened perpendicular to the magnetic field. The corrected mass-to-flux ratio ($\lambda$) becomes 
\begin{equation}\label{eq:mf}
\lambda=\frac{\lambda_{obs}}{3}.
\end{equation}

The resulting mass-to-flux ratios are around 0.7--1.5 for all regions in SDC13. 
This suggests that both the hub and the filament regions in SDC13 are about transcritical. Different correction factors, rather than the factor $1/3$ used in \autoref{eq:mf}, are suggested for different cloud and magnetic field geometries, e.g., $\pi/4$ for a spherical cloud \citep{cr04}, and $3/4$ for a prolate spheroid elongated along the magnetic field \citep{pl16}. Adopting these numerical factors, the mass-to-flux ratios become trans- to supercritical with values ranging from 1 to 3.
We emphasize that the field strengths and mass-to-flux ratios derived here are based on values {\it averaged} over the selected regions. Therefore, they have to be interpreted as {\it average global properties}.
{\it Local higher-density} regions, within regions with a criticality of 0.5 to 3, are likely (highly) supercritical, consistent with the presence of star-formation activity.


\subsubsection{Skalidis \& Tassis Method}\label{sec:ST}
\citet{sk21} point out that observational results show that turbulence in the ISM is anisotropic and that non-Alfv\'{e}nic (compressible) modes may be important. Hence, they propose a new method to estimate the magnetic field strength in the ISM considering these compressible modes, leading to
\begin{equation}\label{eq:ST}
B_{pos}=\sqrt[]{2\pi \rho}\frac{\sigma_{v,NT}}{\sqrt{\delta \phi}}.
\end{equation}
The resulting magnetic field strengths and the corresponding mass-to-flux ratios using the ST method are listed in \autoref{tab:CF}. Field strengths estimated from the ST method are similar to the ones derived from the DCF method within about 30\%. The mass-to-flux ratios are mostly around unity.

\subsubsection{Virial Analysis}\label{sec:Virial}
In this section, we apply the virial theorem to evaluate the relative importance between gravitational, magnetic, and kinetic energy in the central hub. Since the gravitational potential and pressure of a cylinder follow a formalism different from the one for a sphere, the virial analysis for the filament regions is presented in \autoref{sec:fi_stab}. In Lagrangian form, the virial theorem can be written as
\begin{equation}\label{eq:virial}
\frac{1}{2}\ddot{I}=2(\mathcal{T}-\mathcal{T}_s)+\mathcal{M}+\mathcal{W},
\end{equation}
\citep[e.g.,][]{me56,mc07} where $I$ is a quantity proportional to the trace of the inertia tensor of a cloud. The sign of $\ddot{I}$ determines the acceleration of the expansion or contraction of the spherical cloud. The term
\begin{equation}
\mathcal{T}=\frac{3}{2}M\sigma^2_{obs}
\end{equation}
is the total kinetic energy, where M is the total mass and $\sigma_{obs}$ is the observed total velocity dispersion. We neglect the surface kinetic term $\mathcal{T}_s$ because we aim to study the self-stability of each region. Nevertheless, we note that the presence of any external pressure could suppress the support and enhance the cloud's instability. The magnetic energy term, without any force from an external magnetic field, is 
\begin{equation}
\mathcal{M}=\frac{1}{2}Mv^2_A,
\end{equation}
where $v_A=B/\sqrt{4\pi\rho}$ is the Alfv\'{e}n velocity
and $\rho$ is the mean density. We note that the magnetic field morphology is not explicitly accounted for in this magnetic energy term. As the DCF and SF method only constrain the plane-of-sky magnetic field component, we use the statistical average to correct and estimate the total magnetic field strength as $B = (4/\pi)B_{pos}$ \citep{cr04}. The term
\begin{equation}
\mathcal{W}=-\frac{3}{5}\frac{GM^2}{R}
\end{equation}
is the gravitational potential of a sphere with a uniform density $\rho$ and a radius $R$. 

The resulting energy ratios for the central hub in SDC13 are a kinetic-to-gravitational energy $\abs{\mathcal{T}/\mathcal{W}}=0.04\pm0.02$ and a magnetic-to-gravitational energy $\abs{\mathcal{M}/\mathcal{W}}=0.05\pm0.01$. This suggests that {\it globally} gravity is dominating over both kinetic and magnetic energy in the central hub. Moreover, the magnetic energy is comparably important as the kinetic energy, with a derived Alfv\'{e}nic Mach number of $0.6\pm0.1$. The combined ratio $\abs{\frac{2\mathcal{T+M}}{\mathcal{W}}}$ is $0.09\pm0.02$, smaller than unity, suggesting that even the combined kinetic and magnetic energy cannot support the system, and hence the central hub is contracting globally.

\begin{deluxetable*}{ccccccccc}
\tablecaption{Magnetic field strengths and mass-to-flux ratios in SDC13.\label{tab:CF}}
\renewcommand{\thetable}{\arabic{table}}
\tablenum{2}
\tablehead{\colhead{Regions} & \colhead{$n_{H_2}$} & \colhead{$N_{H_2}$} & \colhead{$\sigma_{v,NT}$} & \colhead{$\delta \phi$} & \colhead{$B_{pos}$(DCF)} & \colhead{$\lambda$ (DCF)} & \colhead{$B_{pos}$(ST)} & \colhead{$\lambda$ (ST)}\\
\colhead{} & \colhead{$\textrm{cm}^{-3}$} &\colhead{$\textrm{cm}^{-2}$} &\colhead{(\kms)} &\colhead{(deg)} &\colhead{($\mu$G)} & \colhead{} & \colhead{($\mu$G)} & \colhead{}}
\startdata
\hline
Hub & $(4.2\pm0.4)\times10^4$ & $(3.2\pm0.3)\times10^{22}$ & $0.41\pm0.05$ & $18.5\pm1.0$ & $94\pm5$ & $0.87\pm0.05$ & $75\pm2$ & $1.08\pm0.03$\\
Filament NE & $(2.5\pm0.2)\times10^4$ & $(1.8\pm0.1)\times10^{22}$ & $0.32\pm0.05$ & $35.5\pm1.4$ & $31\pm1$ & $1.50\pm0.05$ & $34\pm1$ & $1.34\pm0.03$\\
Filament NW & $(1.1\pm0.2)\times10^4$ & $(1.3\pm0.1)\times10^{22}$ & $0.22\pm0.05$ & $15.5\pm2.3$ & $34\pm5$ & $0.95\pm0.14$ & $25\pm2$ & $1.29\pm0.10$\\
Filament S & $(2.9\pm0.3)\times10^4$ & $(1.5\pm1.0)\times10^{22}$ & $0.32\pm0.05$ & $20.2\pm1.2$ & $58\pm4$ & $0.66\pm0.03$ & $49\pm1$ & $0.79\pm0.02$\\
\enddata
\tablecomments{Magnetic field strengths and mass-to-flux ratios derived from the DCF and ST method. The uncertainties listed here are obtained from propagating the observational uncertainties through the corresponding equations. Possible additional systematic uncertainties due to, e.g., the unknown dust opacity, are not included. }
{\addtocounter{table}{-1}}
\end{deluxetable*}

\subsubsection{Stability of Filaments}\label{sec:fi_stab}
\citet{wi18} show that the filaments in SDC13 are {\it on average} thermally supercricital, suggesting that the thermal energy is insufficient to support the filaments against gravitational collapse and fragmentation. In this section, we further investigate how the filament criticality varies spatially, considering the support from thermal, non-thermal, and magnetic energy. Based on the Virial theorem, the stability of filaments is commonly evaluated using the critical line density, (or critical line mass), $M_{line,critical}=2(c_s^2+\sigma_{v,NT}^2)/G$ \citep{fi20}, which considers the balance between gravity and the support from both the thermal and turbulent energy. In order to reveal how the filament stability in SDC13 spatially changes, we use the local column density, filament width, and \nh3 velocity dispersion from the previous sections to estimate the filament criticality pixel by pixel. The local filament line density is calculated as $M_{line}=\Sigma W_{F}$, where $\Sigma = \mu m_{H_2}N_{H_2}$ is the filament peak surface density assuming a mean molecular weight $\mu$ of 2.33, and $W_{F}$ is the deconvolved filament FWHM width. The filament criticality is defined as $M_{line}/M_{line,critical}$. Hence, a criticality larger than unity is suggestive of a collapsing and fragmenting filament.

\begin{figure}
\includegraphics[width=\columnwidth]{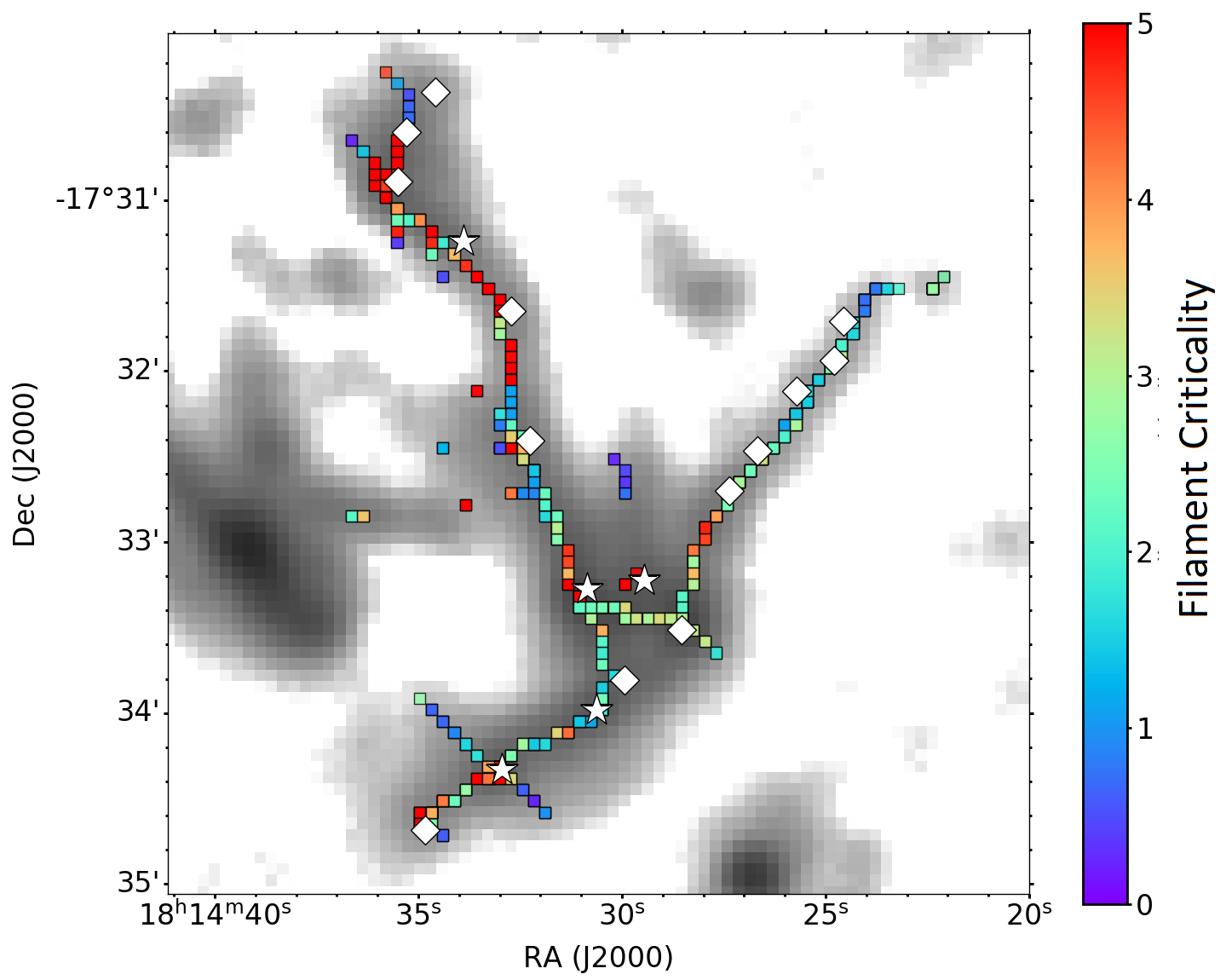}
\caption{Filament criticality in SDC13 overlaid on the JCMT 850 $\mu$m continuum map. The color of the squares is the estimated local filament criticality, considering the support from the thermal and non-thermal energy. The white diamonds and stars label the starless and protostellar sources identified in \citet{pe14}. The filaments are in most locations supercritical, with only a few subcritical locations typically at the tips of the filaments in the outer diffuse zones. If considering the support from magnetic fields, the filament criticality can change within a range from $\pm$5\%, depending on the magnetic field geometry. }\label{fig:fi_cri}
\end{figure}

The calculated filament criticality is visualized in \autoref{fig:fi_cri}. We generally find that most filaments in SDC13 are supercritical. This is consistent with the presence of the numerous starless and protostellar cores along the filaments (\autoref{fig:Fmap}). A few filaments show  subcritical locations in the outer diffuse areas. These are typically at the tips of the filaments. Around the dense cores in the filament NE and S, the local filament criticality is clearly increasing from subcritical in the outer areas to supercritical in the dense cores. Since the change in filament criticality is mainly due to the increasing local column density, this likely indicates that these filaments transition from sub- to supercritical via accumulating mass, possibly through  converging filaments, which then destabilizes the filaments. Subsequently, this will trigger core fragmentation and (future) star formation, as witnessed by the presence of protostellar and starless cores.


Additionally to the support from gas kinetic energy, including thermal and non-thermal, magnetic fields can play a role in stabilizing filaments. However, the exact magnetic support depends on the morphology of the magnetic field within a filament, for which our current data do not have sufficient resolution. \citet{fi20} model magnetized filaments based on the virial theory and suggest that poloidal-dominated fields help supporting filaments against gravity while toroidal-dominated fields destabilize filaments. The critical linear density of a magnetized filament is $M^{mag}_{line,vir}=M_{line,critical}\times(1\pm\mathcal{M}/\mathcal{W})^{-1}$, where $\mathcal{M}$ is the magnetic energy per unit length (with a positive sign for poloidal fields and a negative sign for toroidal fields), and $\mathcal{W}$ is the gravitational energy per unit length. Adopting a $\abs{\mathcal{M}/\mathcal{W}}$ of $0.05\pm0.01$, as derived in \autoref{sec:Virial}, the filament criticality can change within $\pm$5\%.
Hence, even if the magnetic field is poloidal-dominated, the filaments around the dense cores likely remain supercritical, and only small stretches of the filament's outer diffuse regions might transition from supercritical to subcritical. Therefore, the overall finding of the filaments generally being supercritical in SDC13 remains valid even in the presence of magnetic fields, confirming that the conclusion in \citet{wi18} is still valid even considering the magnetic and turbulent energy.

Finally, we note that the underlying assumption to estimate the filament criticality is that the observed velocity dispersion trace the internal turbulent motions. This assumption might not be fully valid within the central hub, where multiple filaments and cores partially overlap, and resulting complex velocity structures might all contribute to the observed velocity dispersion. {\citet{wi18} suggest that the increasing velocity dispersion toward the dense regions in SDC13 is possibly driven by the gravitational fragmentation and infall motion. Such motion is guided by the local gravity and does not support the filaments as efficiently as isotropic turbulence.} This is possibly the reason why the growing filament criticality, as seen toward the dense cores in filament NE and S, is not that evident in the central hub. 

\section{Discussion}\label{sec:dis}
\subsection{Large-scale Environment of SDC13}\label{sec:env10pc}
The pc-scale arc-like magnetic field structure in our polarization map (\autoref{fig:Bmap_SDC13}), distinctively different from the overall magnetic field morphology in SDC13, 
suggests an impact from a cloud-scale event.
Although core formation/fragmentation and star-formation activities can also possibly modify the magnetic field structures \citep[e.g.,][]{zh14,ko14,li15}, these events can typically influence the local magnetic field at $\lesssim$0.1 pc-scales. They are, thus, not likely to explain pc-scale arc-like structures. In order to investigate the possible origin of the arc-like structure in SDC13, this section aims at exploring the 10-pc larger-scale surrounding environment using archival data.

\subsubsection{Magnetic Field Probed by \textit{PLANCK} 353 GHz Polarization Data}\label{sec:envB_Planck}
\begin{figure*}
\includegraphics[width=\textwidth]{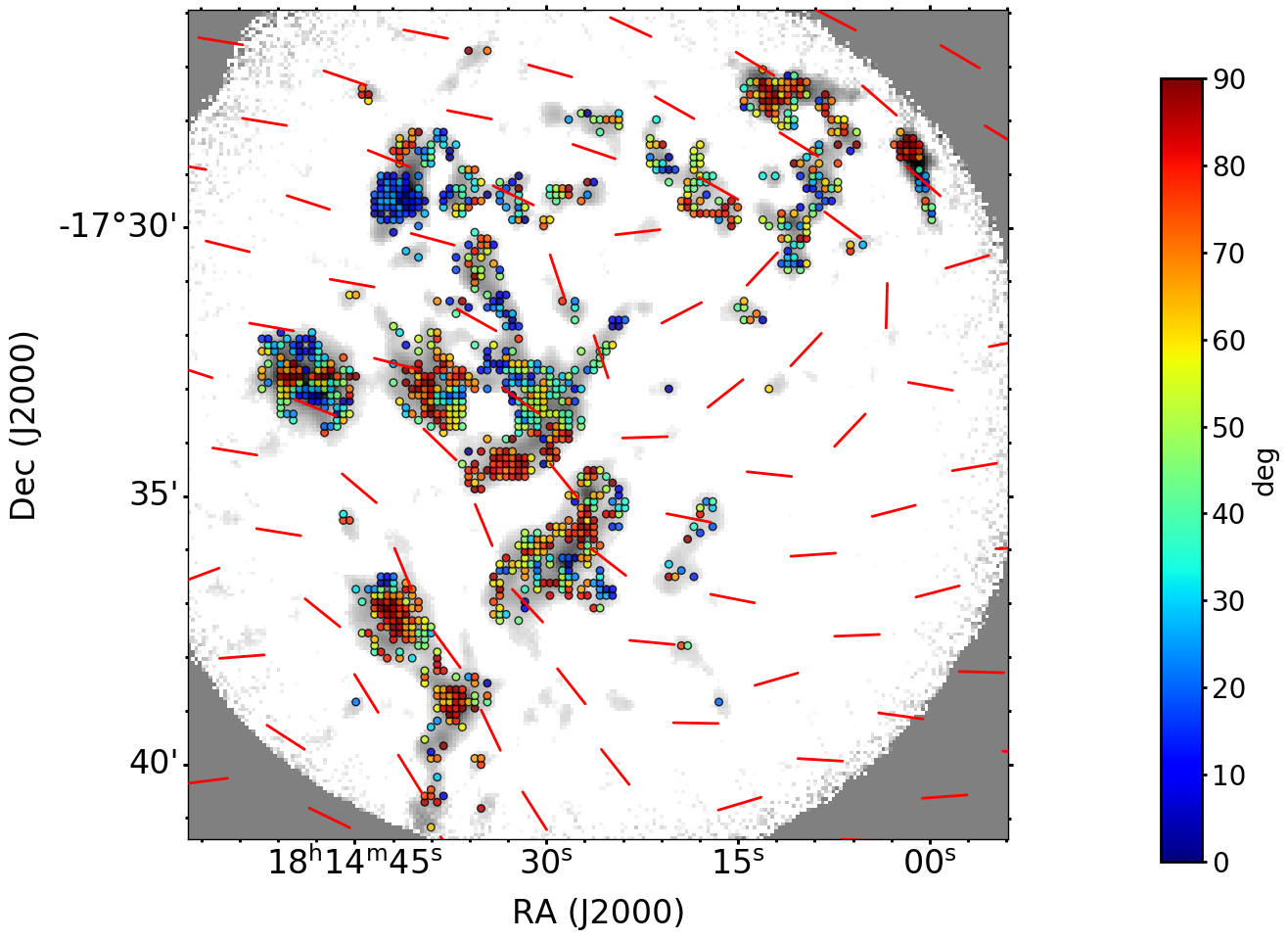}
\caption{Large-scale magnetic field (red segments) traced by \textit{PLANCK} 353 GHz continuum polarization. The \textit{PLANCK} polarization data have a beam size of 5\arcmin, and the segments are plotted per 1/3 beam. The filled colored points represent the relative orientations between the large- and small-scale magnetic field. The large pc-scale magnetic field reveals a converging pattern pointing toward SDC13. The relative orientations show a spatially asymmetric distribution with respect to SDC13,
i.e., the small-scale magnetic field is perpendicular to the large-scale magnetic field in the south of SDC13, and tends to be more parallel to the large-scale magnetic field in the north.}\label{fig:PLANCK}
\end{figure*}

In order to reveal the 10-pc scale magnetic field, \autoref{fig:PLANCK} shows the 353 GHz \textit{PLANCK} magnetic field segments, with a beam size of 5\arcmin, overlaid on our JCMT POL-2 polarization data. The large-scale magnetic field traced by \textit{PLANCK} reveals a partially spiraling and converging morphology, possibly also resembling two incoming wings from the east and west, pointing toward the center of SDC13. This large-scale converging magnetic field pattern tends to locally align with the filament NE and to be perpendicular to filament S and NW. Similarly, the large-scale magnetic field tends to be either parallel or perpendicular to the nearby compact clouds detected in our POL-2 map (also see \autoref{sec:compact_clouds}). 

Magnetic fields either parallel or perpendicular to filamentary clouds have been found from statistics of \textit{PLANCK} polarization \citep{pl16} and starlight polarization data \citep{li13}.
A number of models have been proposed to explain the origin of this configuration, such as magnetic-field-channelled turbulence/shock compression \citep{na08,in09a,in09b,ch20} or filament-filament collision \citep{na14}. In these models, magnetic fields are important in channeling and guiding the mass accretion and subsequent cloud collapse. Following these scenarios, the observed large-scale converging magnetic field morphology in \autoref{fig:PLANCK} is possibly guiding the large-scale gas flows converging towards SDC13. A similarly converging magnetic field, also guiding converging accretion flows, has been seen in another hub-filament system, G33.92+0.11 \citep{wa21}, although at a smaller parsec scale.


In order to reveal how the magnetic field is impacted by the possible converging events,
we calculate the relative orientations between the small- and large-scale magnetic fields traced by \textit{PLANCK} and POL-2 as displayed in \autoref{fig:PLANCK} with the colored points. We find a trend that the relative orientations vary from $\sim0\degr$ to $\sim90\degr$ from north to south within SDC13. 
Similar variations also occur in the nearby compact clouds where the small-scale magnetic field is similarly oriented as the large-scale magnetic field on one side, but then differs in its orientation on the other side of the cloud. Since these large changes in relative orientations occur smoothly over the entire pc-scale cloud, it is unlikely that they originate from the smaller-scale 0.1--0.001-pc core fragmentation or related star-formation processes.

The above discussed variation from large- to small-scale magnetic field morphology can be explained by a converging flow scenario. The induced shocks at the flow colliding layers could compress the local magnetic field, and lead to a magnetic field parallel to the compressing layer or perpendicular to the converging flow \citep{pe12,in18}. In SDC13, the large-relative-orientation areas in \autoref{fig:PLANCK}, covering the filament S and the southern part of the central hub, are perpendicular to the large-scale magnetic field. This is consistent with the expectation if the southern part of SDC13 originates from a flow collision guided by the large-scale magnetic field. In the nearby compact clouds, the large-relative-orientation areas are commonly either parallel or perpendicular to the large-scale magnetic field (see \autoref{sec:compact_clouds} for a brief description on the individual compact clouds).

\subsubsection{Large-scale Gravitational Field Inferred from \textit{Herschel} Continuum Data}
\begin{figure}
\includegraphics[width=\columnwidth]{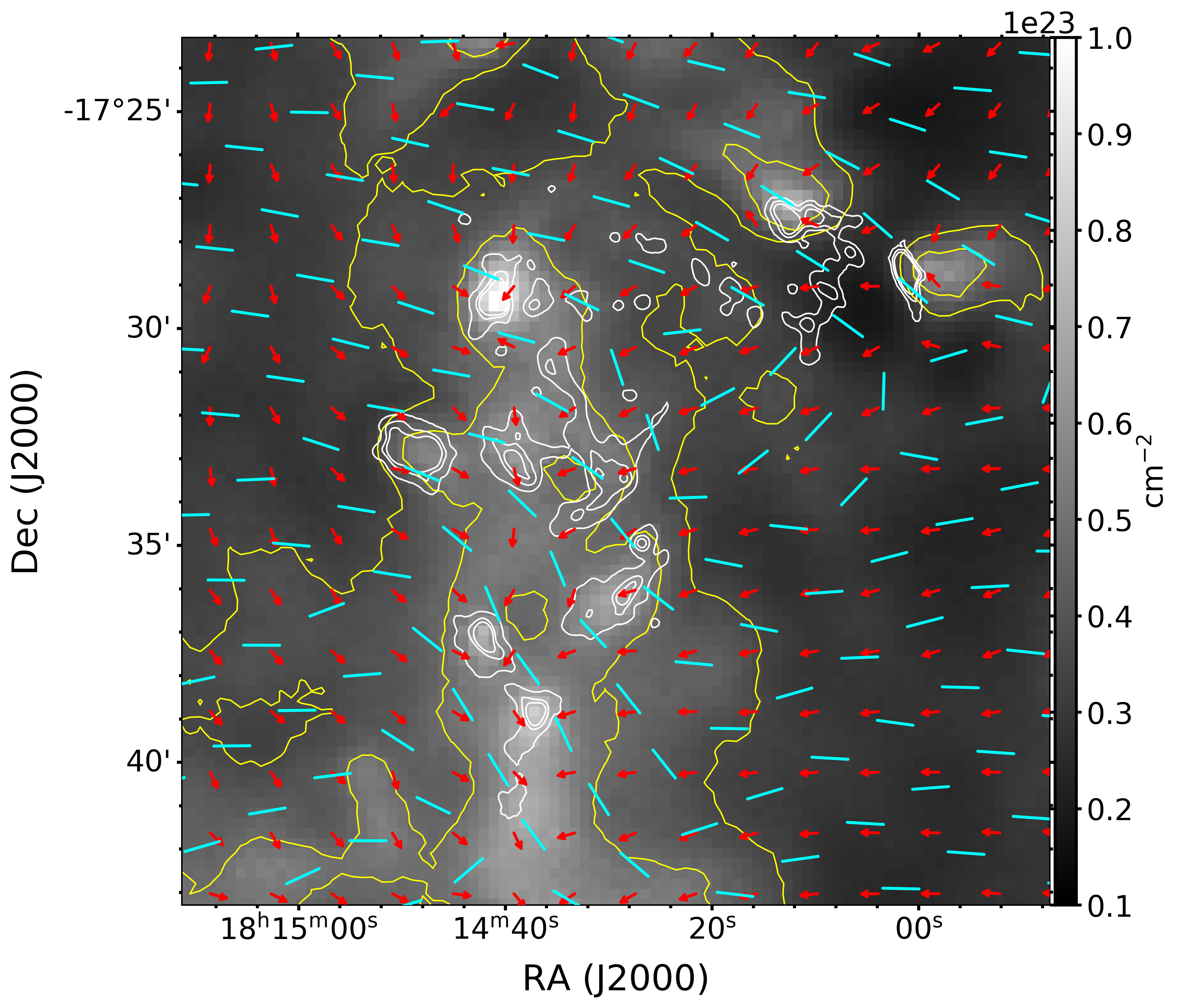}
\caption{Large-scale gravitational field (red arrows) and magnetic field (cyan segments from \textit{PLANCK}) overlaid on the column density map derived from $Herschel$ 5-band continuum data. The white contours are the JCMT 850 $\mu$m continuum as shown in \autoref{fig:BDISmap_SDC13}. The yellow contours mark column density levels of 3.5 and 4.8$\times10^{22} \textrm{cm}^{-2}$, tracing the outer large-scale north-south filament that is hosting SDC13. The large-scale gravitational field is first converging onto the large-scale north-south filament in the outer part, and then pointing toward the south along the filament.}\label{fig:herschel}
\end{figure}

In order to recover the large-scale density structures filtered out by the JCMT POL-2 observations, we used the $Herschel$ archival 5-band continuum data within an area of $2\times2$ deg$^2$ around SDC13 to construct a column density map. We smoothed the 70-500 $\mu$m continuum data to a resolution of 35\arcsec, and fit them with a gray-body function using a dust opacity $\kappa$ of 0.012 cm$^2$/g and a $\beta$ of 2 \citep{hi83}. A zoom-in column density map around SDC13 is shown in \autoref{fig:herschel}. 
This map reveals that SDC13 is embedded in a large-scale north-south filament with a width of $\sim8$ pc. The \textit{PLANCK} magnetic field outside of this filament appears to be relatively uniform and mostly perpendicular to this filament, while the converging magnetic field pattern, as described in \autoref{sec:envB_Planck}, is within this filament. This suggests that the converging  pattern is likely linked to and driven by the kinematics of the outer larger-scale filament.

Simulations predict that the formation of filaments threaded by perpendicular magnetic fields can twist the magnetic field and form a toroidal magnetic field morphology \citep[e.g.,][]{li19}. Furthermore, predicted helical magnetic fields have been observed in a number of clouds \citep[e.g.,][]{ta18}. The spiral-like converging magnetic field within the SDC13-encompassing large-scale filament might also be explained by an inclined toroidal or semi-helical magnetic field twisted by the large-scale filament. Adopting such a scenario, this encompassing large-scale filament threaded by the twisted magnetic field might be replenished by gas flows that are in turn guided by the magnetic field to a central converging region. This mechanism has also been suggested by joint gas kinematics and polarization studies toward other large-scale filamentary systems. Striations near filaments have been commonly found converging toward filament crests along plane-of-sky magnetic fields, e.g., Taurus B211/213 \citep{pal13}, Musca \citep{bo20a,bo20b}. Possible helical magnetic fields have also been detected surrounding large-scale filaments, e.g., Orion A, California, and Perseus \citep{ta18}.
Hence, the collision of converging flows might be the origin of the compact clouds embedded within the large-scale encompassing filament.

To test whether the large-scale gravitational field is consistent with the above outlined scenario, we calculate the projected gravitational field using the 2$\times$2 deg$^2$ column density map, following our approach in \autoref{sec:gravity} (\autoref{fig:herschel}). The gravitational field outside of the encompassing large-scale filament appears to first converge towards the filament ridge along the large-scale magnetic field from north-east and west. 
Getting closer to the ridge of the large-scale filament, the gravitational field is turning towards south to become predominantly aligned with the orientation of the large-scale filament. The gravitational field also gradually becomes aligned with the magnetic field as it approaches the filament from the eastern and western side (between the first and second contour).
This is consistent with the above scenario where the magnetic field is channeling the accretion flows. However, we also note that the gravitational field in the diffuse areas, far from the filament ridge, can show an orientation very different from the magnetic field. This might be explained by a magnetic field being more important in binding the gas than gravity in such low-density regions. 
This is possible because the typical mass-to-flux ratios of clouds with densities (N$_{H_2}$) less than 10$^{21}$ cm$^{-2}$ are likely still magnetically subcritical \citep{cr12}.

\begin{figure*}
\includegraphics[width=\textwidth]{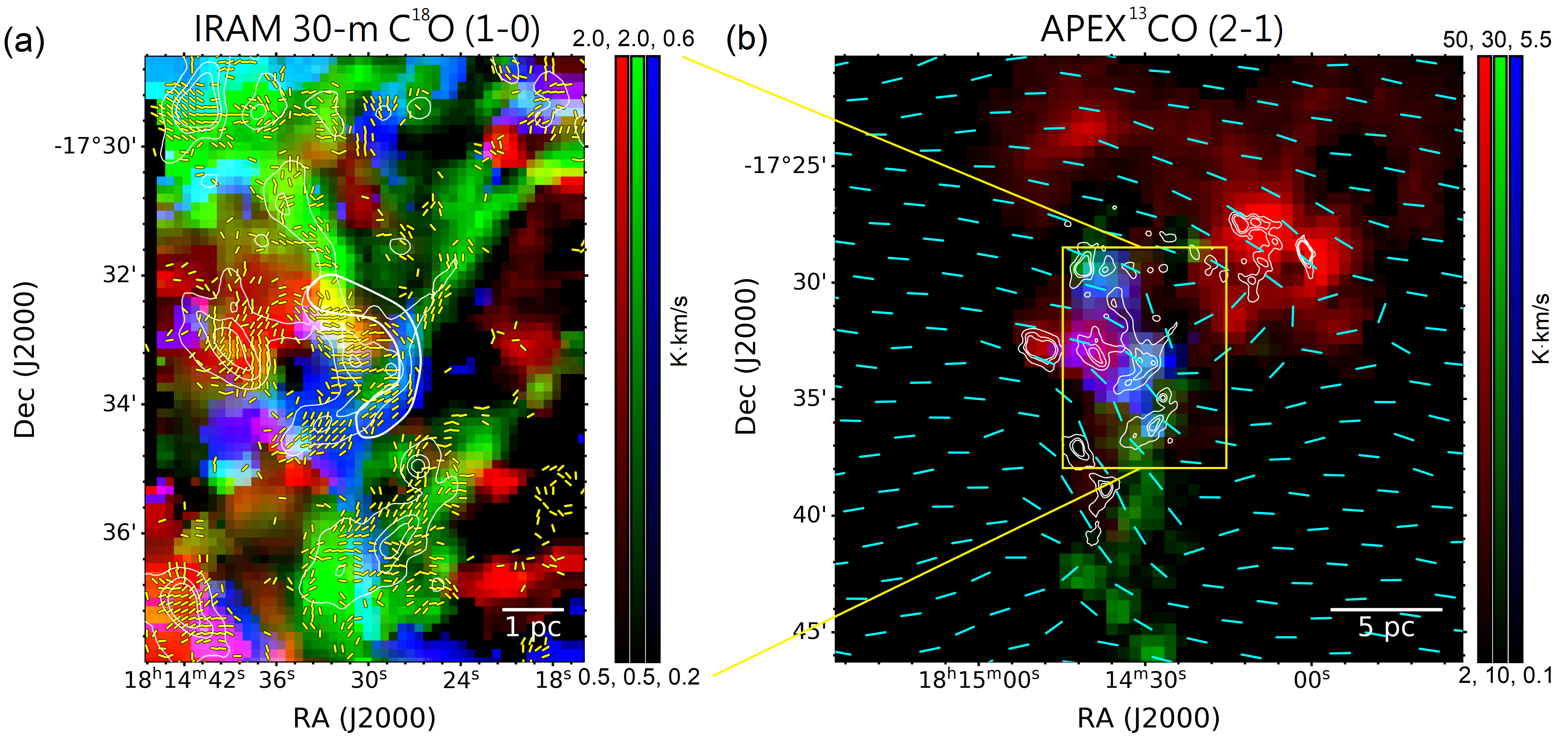}
\caption{(a) Three-color image showing the integrated intensity of the three velocity components, 5--20 km/s (blue), 32--40 km/s(green), and 42--58 km/s (red), associated with SDC13, identified from the IRAM 30-m \textrm{C$^{18}$O} (1-0) data with a resolution of 24.6\arcsec and an rms noise of 0.17 K. The white contours are the 100, 300, and 500 mJy/beam JCMT 850 $\mu$m continuum, and the thick white line marks the arc-like magnetic field structure as shown in \autoref{fig:Bmap_SDC13}. The green component is consistent with the ``Y-shape'' SDC13 HFS. The yellow segments are the larger-than 2$\sigma$ magnetic field segments. The blue component reveals four filamentary arms converging to the clump within the SDC13 filament S, consistent with the filaments identified in the 850 $\mu$m dust continuum. One of these arms extending to the central hub is likely aligned with the arc-like magnetic field structure. The red component connects the eastern to the SDC13 central hub through a bridge. (b) Three-color image showing the three GMCs, SDG013.178-0.0950 (blue), SDG012.840-0.2041 (green), and SDG013.222+0.0076 (red) from the APEX $^{13}$CO (2-1) data with a resolution of 30\arcsec and an rms noise of 1 K, which spatially overlap with SDC13 and which are also associated with the three \textrm{C$^{18}$O} components displayed in the (a). \textbf{The yellow box labels the outer boundary of (a).} SDG012.840-0.2041 is consistent with the large-scale north-south filament seen by $Herschel$ in \autoref{fig:herschel}.
Most of the compact clouds detected in the JCMT 850 $\mu$m continuum are located near the edge or the boundary of these GMCs, either perpendicular or parallel to the large-scale \textit{PLANCK} magnetic field.}\label{fig:GMC}
\end{figure*}

\subsubsection{Nearby Giant Molecular Clouds}
In order to trace the gas possibly flowing along the large-scale magnetic field, we
use IRAM 30-m \textrm{C$^{18}$O} (1-0) data (Williams et al., in prep.) to identify the velocity components that are likely associated with SDC13. This data set (Project ID: 024-13, PI: Peretto) has a beam size of 24.6\arcsec, a spectral resolution of 0.13\kms, and an rms noise of 0.17 K. \autoref{fig:GMC}(a) shows the identified three components, 5--20 km/s (blue), 32--40 km/s (green), and 42--58 km/s (red). The green component clearly displays a ``Y-shape'' morphology, spatially overlapping with the 
"Y-shape'' structure of SDC13 seen in the 850 $\mu$m continuum.
The blue component reveals four filamentary arms converging to the dense core embedded in the filament S. One of these arms is aligned with the arc-like magnetic field pattern and also the eastern boundary of the central hub. The red component is connecting the eastern with the central hub through a bridge.

The spatial arrangement of these \textrm{C$^{18}$O} components supports our earlier outlined scenario where the observed changes from large- to small-scale magnetic field originate from cloud-cloud collisions. The ``U-shape'' morphology of the blue component, aligned with the "U-shape" magnetic field structure, is identical to a shock-compressed layer where the magnetic field is bent by the collision, as predicted in cloud-cloud simulations \citep{in13,in18}. The red component presents a possible incoming flow moving toward SDC13 from the north-east.
Additionally, it probably also acts as a channel from the eastern hub.

To further understand the origin of the \textrm{C$^{18}$O} velocity components, we searched for even larger-scale ``reservoirs''. 
We, therefore, looked for giant molecular clouds (GMCs) identified from APEX $^{13}$CO (2-1) data in the SEDIGISM catalog \citep{sc21,du21}. GMCs were selected based 
on overlapping with SDC13 in position-position-velocity.
To extract only the major cloud components, we excluded GMCs with an angular size smaller than 5\arcmin. The final selected GMCs are SDG013.178-0.0950 (blue), SDG012.840-0.2041 (green), and SDG013.222+0.0076 (red), which have a velocity range of 12--20 km/s, 35--38 km/s, and 47--59 km/s, respectively, consistent with the three velocity components found in the SDC13 IRAM 30-m \textrm{C$^{18}$O} data. \autoref{fig:GMC}(b) presents the integrated intensity maps of these three selected GMCs in different colors. We note that another GMC, SDG013.098-0.0821, also overlaps with SDC13. However, its corresponding \textrm{C$^{18}$O} component, in the range 20--28 \kms, is associated with the compact cloud SDC13.121-0.091 (south of SDC13, see \autoref{fig:Bmap_all}). Hence, we do not further discuss it here.

SDG012.840-0.2041 (green) is consistent with both the large-scale north-south filament revealed by the $Herschel$ column density map, and also with the main structure of SDC13 in the \textrm{C$^{18}$O} data. Its velocity range is consistent with the \nh3 and \textrm{C$^{18}$O} velocity component tracing the main structure of SDC13. SDG013.178-0.0950 (blue) and SDG013.222+0.0076 (red) appear with an extended morphology likely winding around the north-south filament along the magnetic field. The velocity difference between these two GMCs and SDG012.840-0.2041 is about 10--20 km/s, which is considered a typical range enabling cloud-cloud collisions and the formation of a massive cluster ($\sim$10--40 \kms, \citet{in18,co19,fu21,do21}).

The positions of the compact clouds detected in our JCMT POL-2 data (\autoref{fig:Bmap_all}) are mostly falling onto the conjunction areas between two or three GMCs. SDC13 is embedded in SDG012.840-0.2041, and possibly compressed by the other two GMCs from the south-western and eastern side. 
The two compact clouds (SDC13.121-0.091 and SDC13.123-0.157) in the south of SDC13 are located at the boundaries of these GMCs, with their major axes along the boundaries. A series of small compact clouds north of SDC13 (SDC13.225-0.004 and SDC13.246-0.081) is located along the extending structures of SDG013.222+0.0076 overlapping with SDG012.840-0.2041, and connecting to the two compact clouds (SDC13.190-0.105, SDC13.198-0.135) east of SDC13. 

In summary, this spatial consistency suggests that the interaction between these GMCs is highly correlated with the formation of the compact clouds. We, however, note that the sensitivity of the molecular line data is not yet sufficient to also detect the even diffuser gas that very likely is present within the velocity gaps between these GMCs. Such bridging structures connecting velocity components in the position-velocity diagrams are commonly considered as evidence of cloud-cloud collision \citep[e.g.,][]{ha92}. Future high-sensitivity observations are needed to understand and possibly model the interaction between these GMCs in more detail.

\subsection{Star-Forming Environment within SDC13}\label{sec:env1pc}
Although SDC13 likely originated from a cloud-cloud collision as discussed in the previous sections, the evolution of SDC13 at the current stage is probably driven by gravity. The reasons for this are: (1) The virial analysis shows that the gravitational energy within the central hub is dominating both magnetic and kinematic energy. (2) The extending filaments are mostly supercritical, even when considering additional support from kinematic and magnetic energy. (3) \citet{wi18} found that the \nh3 line width in SDC13 is increasing toward the central hub.
Their analysis suggests that this increased
kinetic energy likely results from a conversion of gravitational energy.
Hence, even though the cloud-cloud collision likely was important in triggering the formation of SDC13 in an early stage, the corresponding large-scale kinematic energy has likely dissipated, leaving smaller-scale dynamical (gravitational) processes to take over in the current stage. 

This gravity-driven picture is further supported by the emerging trends
between the local orientations among gravity, filament, and velocity gradient. \autoref{sec:ana_local} shows that local gravity is pointing toward both the filament ridges and the central hub, causing an offset angle of 20--60\degr in orientations between filaments and local gravity. 
Material within filaments is likely being pulled by local gravity. And, indeed, a similar offset angle of 25--75\degr is observed between filaments and local velocity gradients. This neither parallel (0\degr) nor perpendicular (90\degr) offset angle possibly results from the combination of the two gravitational pulling modes as described in \autoref{sec:gravity}, i.e.,  an offset angle of 90\degr\ is expected if the gas is pulled to the filament ridge, while an offset angle of 0\degr\ is expected if the gas is converging to the center. The range of the observed offset angles might suggest that the two modes are present simultaneously and are comparably important. 
The filaments in SDC13 might accumulate mass via the above modes, and gradually increase their line density. \autoref{fig:fi_cri} shows that the filaments in SDC13 are mostly supercritical, suggesting that gas within these filaments can fragment and form stars {\it before} reaching the central hub. This can explain the presence of protostellar and starless cores along the filaments.

Although the evolution of SDC13 is mainly driven by gravity, the magnetic field might still play some role in shaping the filamentary network in the hub center. \autoref{fig:hist_2I} shows that the filament orientations are parallel to the magnetic field in the low-density regions, but become perpendicular in the high-density regions. This is similar to the change in filament-magnetic field alignment discovered in \citet{pl16} with a transition density of $10^{21.7}$ cm$^{-2}$, which is close to ours (250 \mjyb$\sim3.5\times10^{22}$ cm$^{-2}$). This transition suggests that the role of the magnetic field might evolve with local density also in a hub-filament system. 

\begin{figure}
\includegraphics[width=\columnwidth]{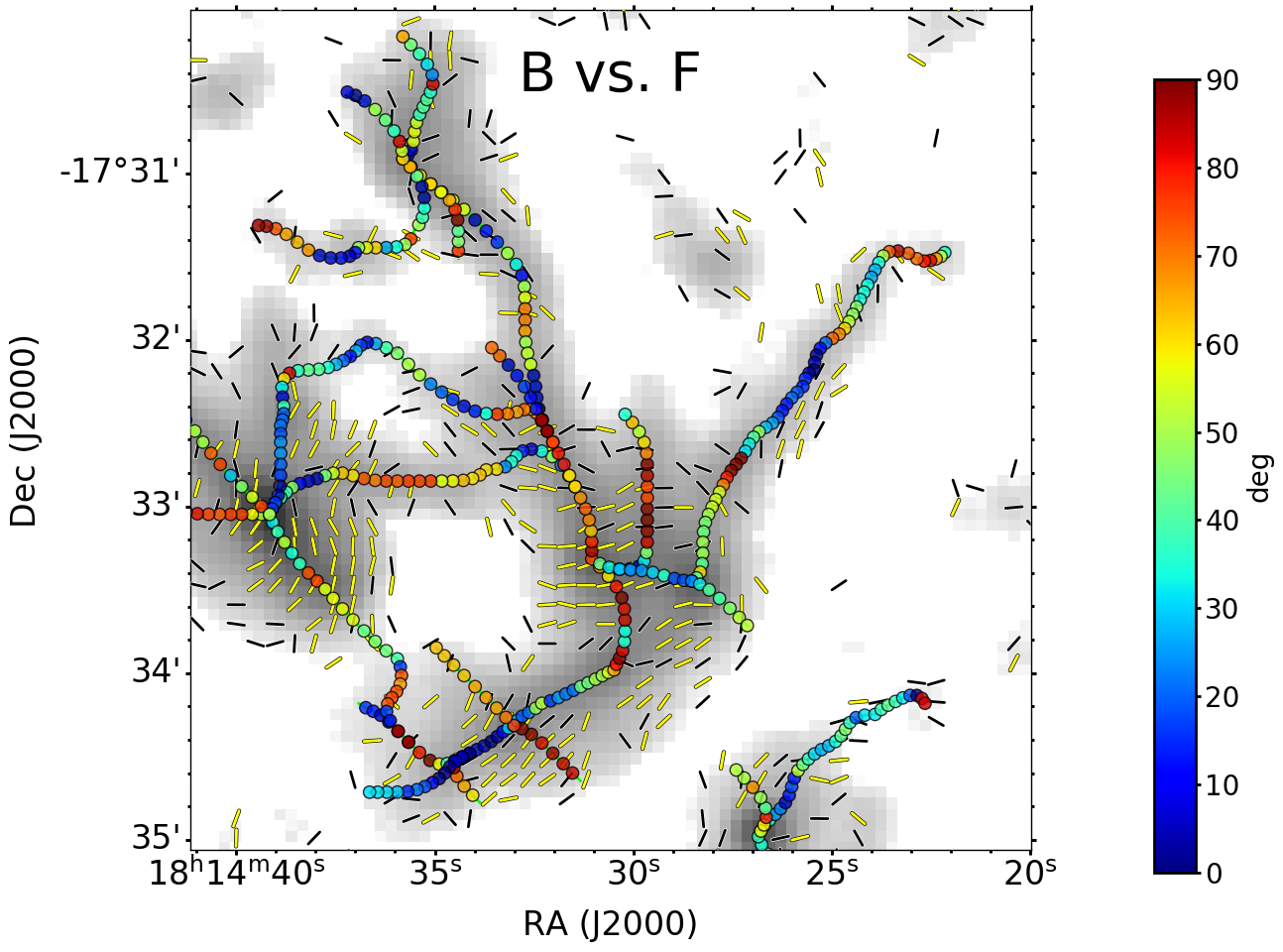}
\caption{ Relative orientations between filaments and magnetic field (color scale), overlaid on the 850 $\mu$m intensity. 
The yellow and black segments are the magnetic field segments as shown in \autoref{fig:Bmap_SDC13}. Within the central hub, the longer filaments tend to be either parallel or perpendicular to the local magnetic fields, except for the filaments near the "U-shape" magnetic field pattern where the field is more complex.}\label{fig:BvsF}
\end{figure}

Indeed, \autoref{sec:ana_local} shows that, within SDC13, local velocity gradients are only correlated with local gravity in low-density regions, and only correlated with the magnetic field in high-density regions, suggesting that the magnetic field in the central hub, possibly enhanced by shock compression, becomes important in regulating the gas motion and hub fragmentation.  \citet{va14} simulate cloud fragmentation under relatively strong magnetic fields, and predict that the fragmented filaments would be parallel to each other and perpendicular to the local magnetic field, which is seen in G14.225-0.506 \citep{bu13}. This predicted morphology is similar to the filaments within the SDC13 central hub, either perpendicular or parallel to the east-west magnetic field (\autoref{fig:BvsF}). Only the filaments near the "U-shape" magnetic field region show different offset angles with respect to the  magnetic field. If the central hub is indeed fragmenting under the regulation of the magnetic field, we will expect to find a centrally condensed protocluster with a higher level of mass segregation, because a strong magnetic field can efficiently suppress core fragmentation \citep{he11,my13}. Future high-resolution data resolving the fragmented cores are essential to further examine this possibility.

\subsection{Hub-Filament System Originating from Cloud-Cloud Collision?} \label{sec:origin}
The comparison between the multi-scale magnetic fields traced by the \textit{PLANCK} and the JCMT POL-2 polarization data reveals a partial spiral-like converging magnetic field pattern surrounding SDC13 at a 10-pc scale and a locally uniform field within SDC13 at a 0.5-pc scale. The locally uniform 0.5-pc scale magnetic field is consistent with the 10-pc magnetic field in the northern part of SDC13, but then becomes perpendicular to the large-scale magnetic field in the southern part. This variation of the magnetic field morphology from large to small scale likely carries the imprint of the evolution of the physical conditions from the initial to the current stage. 

\begin{figure*}
\includegraphics[width=\textwidth]{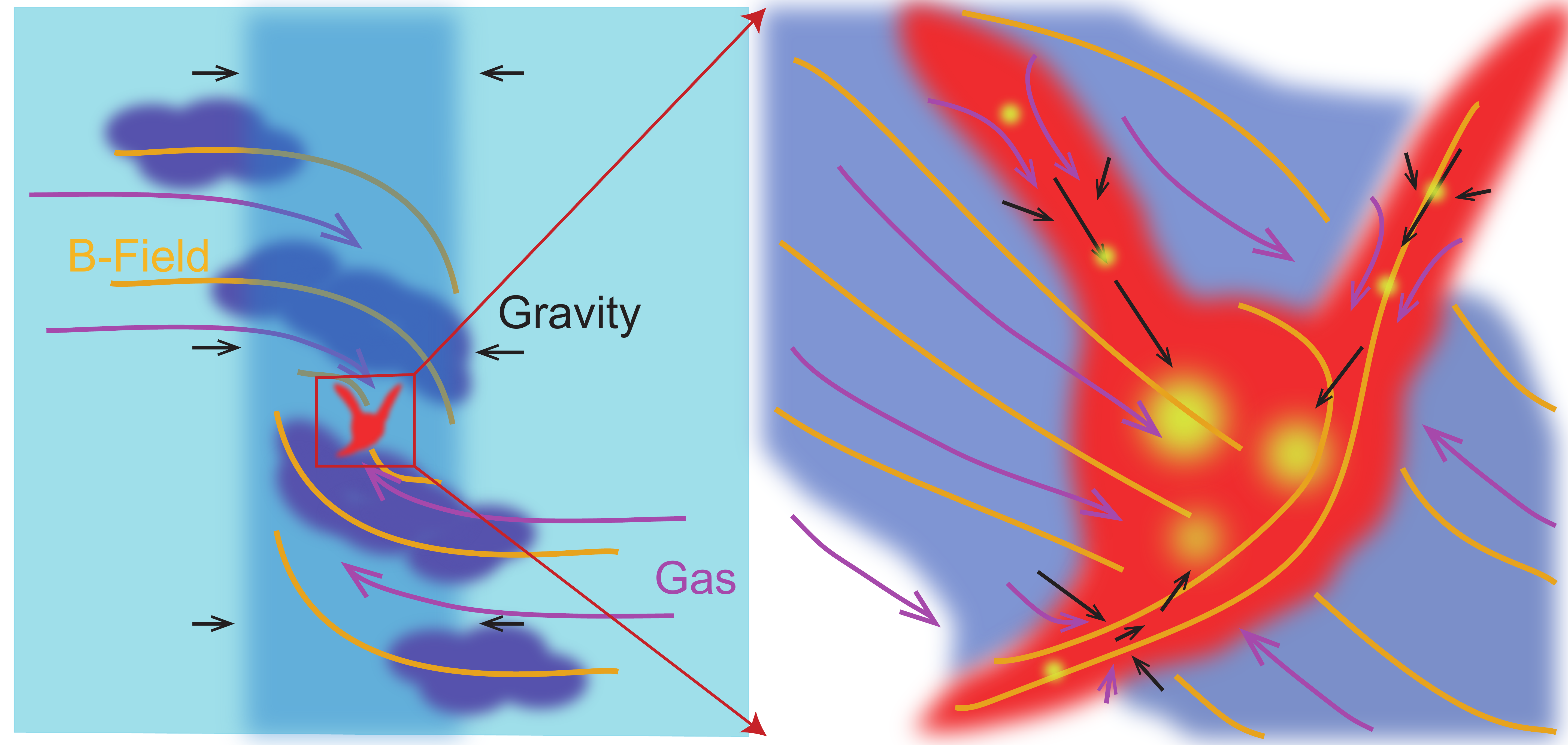}
\caption{Cartoon figure summarizing the various observed features (\autoref{sec:origin}). The left panel illustrates the 10-pc-scale environment around SDC13 (section \ref{sec:env10pc}). The right panel is zooming in to SDC13 (in red; section \ref{sec:env1pc}). Black arrows represent the directions of local gravity. Orange curves visualize the magnetic field morphology. The magenta arrows are the directions of gas motions. The light and darker blue background colors display the different local densities. At the 10-pc scale (left panel), a large-scale filament, 
formed by the collision of large-scale flows with non-zero angular momentum, twists the magnetic field resulting in a large-scale toroidal magnetic field. This toroidal magnetic field guides the ambient denser clouds (dark blue) on their way of being accreted onto the filament. At the pc-scale (right panel), these clouds flow along magnetic field lines and collide at a main convergent point to form a compact hub-filament system with filaments either parallel to the large-scale flow or parallel to the colliding front. At the same time, the small-scale local magnetic field within the shock-compressed layers is enhanced and becomes parallel to the shock-compressed layers and perpendicular to the large-scale magnetic field. After the compact hub-filament system has gained sufficient mass, local gravity starts to dominate over the gas kinematics, leading to gas accretion both toward the filament ridges and the central hub.
As the filaments become mostly supercritical and also the central hub reaches a larger-than-one mass-to-flux ratio in local dense regions, the formation of protostellar and starless cores (yellow regions) is enabled.
}\label{fig:cartoon}
\end{figure*}

Putting together all the pieces, we propose a scenario for the formation and evolution of the SDC13 hub-filament system as illustrated in \autoref{fig:cartoon}. At a 10-100 pc scale, the large-scale north-south filament, traced by the $Herschel$ column density map, is generated from a collision of large-scale MHD flows along magnetic field lines. The angular momentum carried by these colliding flows (with non-zero impact parameters) twist the large-scale magnetic field and enhance its toroidal component \citep{li19}. The ambient material surrounding the large-scale filament keeps feeding the filament through flows along the toroidal magnetic field lines. This leads to the appearance of the observed GMCs winding around the filament ridge.

At pc-scale, the large-scale colliding flows start to converge,
and thus form a compact HFS embedded inside the large-scale filament. The collision and subsequent converging process determine the initial morphology of the resulting HFS with the magnetic field within it. The forming filaments are either aligned with the shock-compressed layers or follow the converging flows. Similarly, the magnetic field is significantly distorted where there are collisions, becoming locally aligned with the compressed layers, while it remains consistent with the large-scale magnetic field outside of the compressed layers. 

While accumulating material through the large-scale flows and after the turbulent energy has dissipated in the post-shock gas at pc-scale, gravity takes over the evolution of this HFS. Gravity both drives the radial contraction of the filaments and pulls the gas towards the center of the gravitational well. The filaments can accumulate mass directly from the ambient gas or through the short converging filaments. They eventually become supercritical and start to fragment and form dense cores. \citet{wi18} estimated an averaged dynamical age of $\sim$5.2 Myr for the SDC13 filaments, based on their current densities and assuming constant accretion rates. This can explain the numerous dense cores predominately along the filament ridge and especially at the locations where multiple filaments converge, as the collapsing timescales of these structures (0.1--0.7 Myr, \citep{wi18}) are well below the filament age. This picture agrees with the scenario proposed in \citet{wi18} where the current evolution of SDC13 is predominantly driven by gravity. Although the magnetic field is not sufficiently strong to support the filaments against gravitational collapse at pc-scale, it can still play a role in regulating the direction of fragmentation. This results in a filamentary network where filaments are parallel to the local magnetic field in the low-density regions, being dragged by the gravitating flows, but then become perpendicular to the magnetic field in the high-density regions, because cloud fragmentation is more efficient along magnetic fields.

We note that \citet{ku20} propose a scenario where HFSs originate from flow-driven filament-filament collisions. Our scenario provides an alternative process to form HFSs. The major difference of these two scenarios is that \citet{ku20} propose a side-by-side collision between gravitationally unbound filaments, driven by the intra-molecular cloud velocity dispersion or expanding shells, while our scenario suggests a head-on collision of gravitationally-driven and magnetically-guided flows converging onto the 10-pc-scale filament (\autoref{fig:herschel}, \autoref{fig:GMC}, \autoref{fig:cartoon}).

\section{Summary}\label{sec:con}
This paper conducts a study of the hub-filament system SDC13 using JCMT SCUBA-2/POL-2 850 $\mu$m continuum polarization observations. Our polarization data reveal an organized but complex magnetic field morphology. From the analysis of these data, 
together with ancillary velocity and additional data covering the larger-scale surrounding of SDC13, we find the following results.

\begin{itemize}
    \item The magnetic field within the dense regions of SDC13 appears to be relatively uniform. However, a cloud-scale arc-like 
    "U-shape"
    magnetic field feature is identified along the 
    western
    boundary of the central hub (\autoref{fig:Bmap_SDC13}). This separate feature has a distinctively larger magnetic field dispersion (\autoref{fig:BDISmap_SDC13}).
    
    \item Filamentary structures are identified in SDC13 using the $DisPerSE$ algorithm. The major filaments form a "Y"-shaped network, with a number of minor filaments converging onto them. 
    Without any exception,
    all known starless and protostellar cores are located along the identified filaments, while the protostellar cores additionally tend to be distributed at the filament convergent points (\autoref{sec:filament}).
    
    \item Local gravity in SDC13 hints at a combination of two modes: a gravitational pull towards filament ridges and a pull towards filament convergent points. Additionally, the directions of local gravity are correlated with the directions of local velocity gradients in the low-density regions.
    All together, this suggests that the gas motions in SDC13 are also following the same two modes: gas is locally accreting onto filament ridges and globally converging to the central hub. These two dominant modes indicate that SDC13 is undergoing a multi-scale gravitational collapse (\autoref{sec:ana_local}).
                    
    \item The local magnetic field in SDC13 tends to be more parallel to filaments in low-density regions and more perpendicular to filaments in high-density regions (\autoref{fig:hist_2I}). This points to a role of the magnetic field than can vary, from channeling accreting gas to regulating cloud fragmentation. 
    Our local analysis determines the filaments to be supercritical in most locations (\autoref{sec:fi_stab}).

    \item {Globally, the magnetic field strengths estimated from the Davis-Chandrasekhar-Fermi method indicate that the mass-to-flux ratios in SDC13 are transcritical to supercritical. A virial analysis, finding that the gravitational energy is larger than the kinematic and magnetic energy, suggests that SDC13 is globally collapsing (\autoref{sec:ana_global}).
    }
    
    \item Comparing to the \textit{PLANCK} large-scale magnetic field, the small-scale  magnetic field in SDC13 is locally aligned with the large-scale magnetic field in the north-eastern side, but locally perpendicular to the large-scale magnetic field in the south-western side where the arc-like magnetic field feature is located. Moreover, the larger-scale $^{13}$CO and \textrm{C$^{18}$O} molecular line data show that these two different magnetic field morphologies might be associated with two neighboring giant molecular clouds, with a velocity difference of 10--20 \kms (\autoref{sec:env10pc}). 

    \item Combining all the findings from both the large-scale and the small-scale environment, we propose a two-stage scenario to explain the formation of SDC13. In a first stage, the large-scale GMCs collide with each other along the large-scale toroidal magnetic field. SDC13 is formed within the shock-compressed layers, where magnetic field and filaments are aligned with the shock front. In a second stage, after the shock kinetic energy has dissipated, local gravity takes over the evolution of the system, driving the mass locally accreting onto the filaments and globally converging to the central hub. Protostellar cores can then form after the filaments or convergent points have accumulated sufficient mass (\autoref{fig:cartoon}).
    
\end{itemize}

\acknowledgments
The James Clerk Maxwell Telescope is operated by the East Asian Observatory on behalf of The National Astronomical Observatory of Japan; Academia Sinica Institute of Astronomy and Astrophysics; the Korea Astronomy and Space Science Institute; the National Astronomical Research Institute of Thailand; Center for Astronomical Mega-Science (as well as the National Key R{\&}D Program of China with No. 2017YFA0402700). Additional funding support is provided by the Science and Technology Facilities Council of the United Kingdom and participating universities and organizations in the United Kingdom and Canada. Additional funds for the construction of SCUBA-2 were provided by the Canada Foundation for Innovation. The authors wish to recognize and acknowledge the very significant cultural role and reverence that the summit of Maunakea has always had within the indigenous Hawaiian community.  We are most fortunate to have the opportunity to conduct observations from this mountain. 
This publication is based on data acquired with the Atacama Pathfinder Experiment (APEX) under programs 092.F-9315 and 193.C-0584. APEX is a collaboration among the Max-Planck-Institute for Radio astronomy, the European Southern Observatory, and the Onsala Space Observatory. The processed data products are available from the SEDIGISM survey database located at https://sedigism.mpifr-bonn.mpg.de/index.html, which was constructed by James Urquhart and hosted by the Max Planck Institute for Radio Astronomy.
PMK is supported by the Ministry of Science and Technology (MoST) through grants MoST 109-2112-M-001-022 
and MoST 110-2112-M-001-057. YWT acknowledges support from MoST 108-2112-M-001-004-MY2. G.A.F acknowledges support from the Collaborative Research Centre 956, funded by the Deutsche Forschungsgemeinschaft (DFG) project ID 184018867.

\appendix
\section{Polarization Properties}\label{sec:pp}
\begin{figure*}
\includegraphics[width=\textwidth]{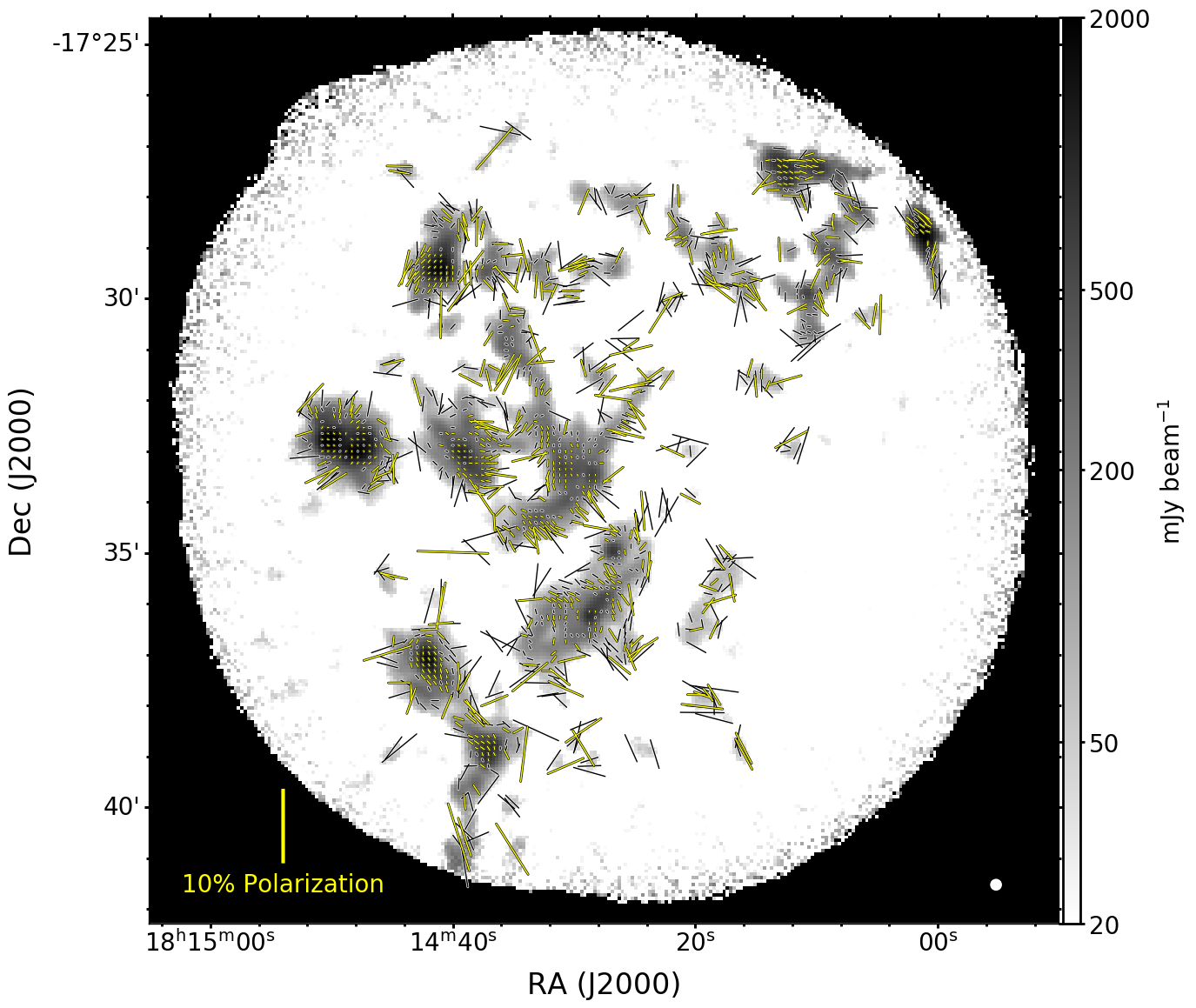}
\caption{POL-2 polarization segments sampled on a 7\arcsec\ grid overlaid on 850 $\mu$m dust continuum, sampled on a 4\arcsec\ grid, of the entire SDC13 region with nearby compact clouds. The yellow and 
black segments display the larger than 3$\sigma$ and 2--3$\sigma$ polarization detections. The lengths of the segments are proportional to the polarization fractions. The yellow scale bar at the left bottom shows a 10 \% polarization fraction. The white circle at the right bottom corner is the JCMT beam size of 14\arcsec. The rms noise of Stokes I is $\sim$1--5 m\jyb\, depending on the pixel intensities, within the central 3\arcmin\ area. 
}\label{fig:pmap_raw}
\end{figure*}

\begin{figure}
\includegraphics[width=\columnwidth]{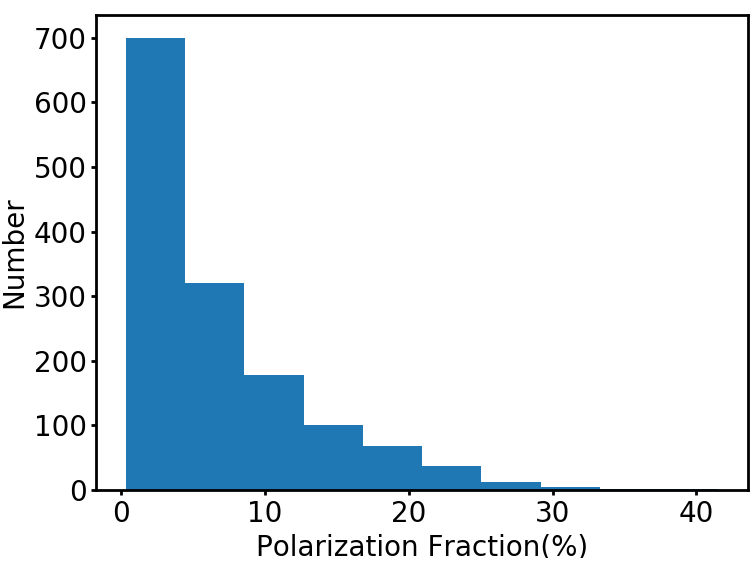}
\caption{Histogram of debiased polarization fraction of the selected $P/\sigma_{P} >2$ samples.
Most of the samples show a polarization fraction smaller than 20\%, and the median polarization fraction is 4.6\%.}\label{fig:P_hist}
\end{figure}

\autoref{fig:pmap_raw} shows the observed POL-2 polarization segments overlaid on the 850 $\mu$m total intensity where the lengths of the segments are proportional to the debiased polarization fraction. 
A histogram of the debiased polarization fraction is presented in \autoref{fig:P_hist}. The median fraction is 4.6\%, and most of the samples have a fraction smaller than 20\%.
The observed polarization fraction is clearly higher near the outskirts of the clouds.
Assuming that dust grains are aligned with the magnetic field is a fundamental assumption that allows us to use polarization data to trace magnetic field structures. 
A common way to examine this assumption is to investigate how the polarization fraction correlates with the total intensity $I$.
A polarization fraction $P$ decreasing with total intensity, following 
$P=PI/I\propto I^{-1}$, is expected if the dust grains are not aligned with the magnetic field in a dense cloud such that the polarized intensity $PI$ is independent of the total intensity. In contrast, if magnetic-field-aligned dust grains are present in a dense cloud, $PI$ would increase with the cloud's column density, which results in a $P-I$ relation where $P\propto I^{-\alpha}$ with $\alpha$ smaller than unity.

\begin{figure}
\includegraphics[width=\columnwidth]{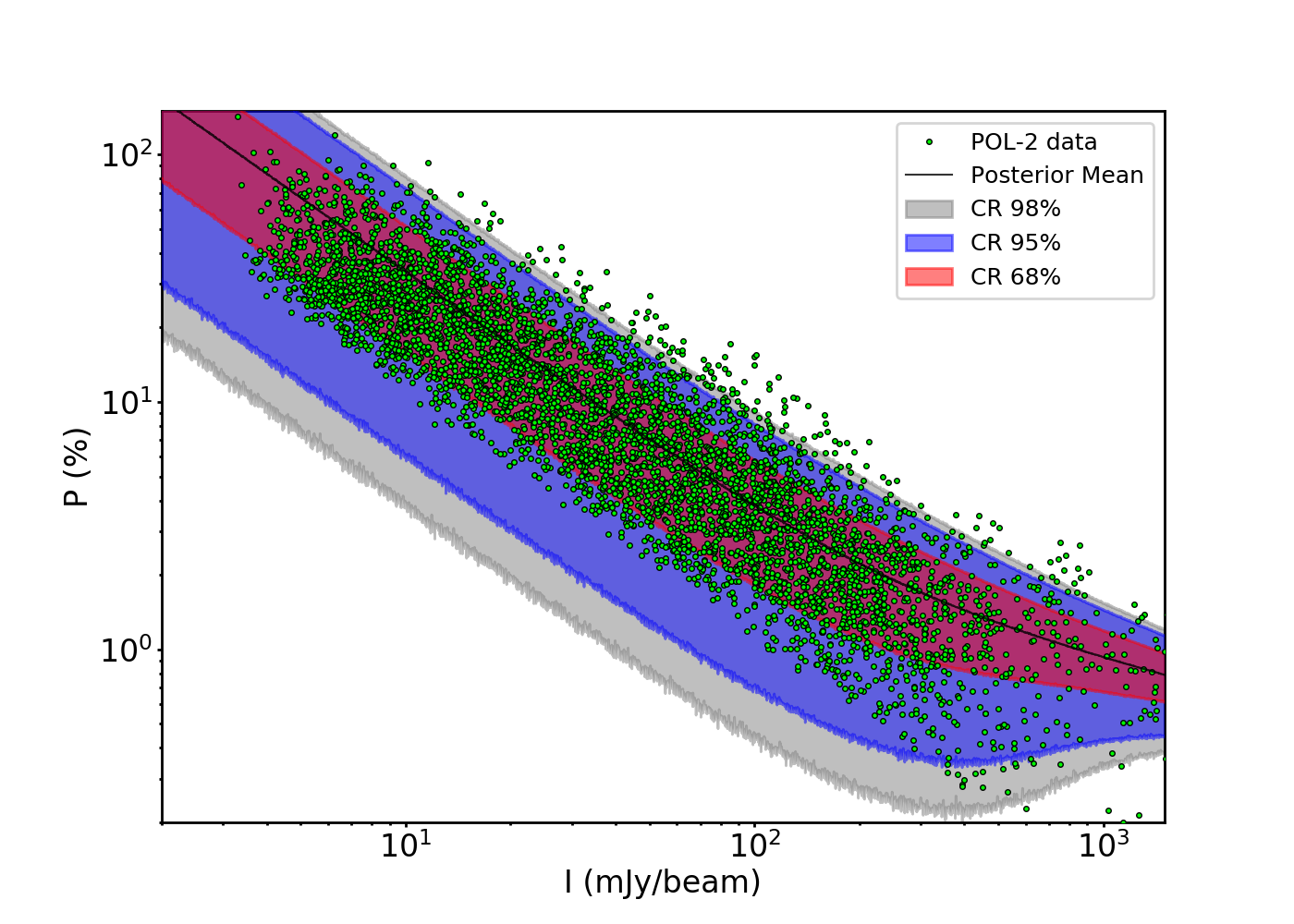}
\caption{850 $\mu$m total intensity $I$ vs polarization fraction $P$. The green points are the non-debiased, $I/\sigma_{I}>$3, and $\sigma_I<$5 m\jyb\ POL-2 polarization measurements. The colored regions are the predicted $I-P$ distributions based on the Bayesian analysis within the 68\%, 95\%, and 98\% confidence regions (CR). The black line indicates the posterior mean. Most of the data points are within the 98\% confidence region of our prediction.}\label{fig:IP}
\end{figure}

The observed $I-P$ relation for SDC13 is displayed in \autoref{fig:IP}. We follow the Bayesian analysis in \citet{wa19} to determine the power-law index $\alpha$ using the model 
\begin{equation}\label{eq:prior}
P =\beta I^{-\alpha}
\end{equation}
with a probability distribution function (PDF) of P described by the Rice distribution
\begin{equation}\label{eq:rice}
F(P|P_0)=\frac{P}{\sigma_P^{2}}\exp\left[-\frac{P^2+P_0^2}{2{\sigma_P}^2}\right]I_0\left(\frac{PP_0}{\sigma_P^2}\right),
\end{equation}
where $P$ is the observed polarization fraction, $P_0$ is the real polarization fraction, $\sigma_P$ is the Ricean dispersion in the polarization fraction, and $I_0$ is the zeroth-order modified Bessel function. We further assume that the uncertainty in the polarization fraction is given by
\begin{equation}
\sigma_P=\sigma/I,
\end{equation}
where $\sigma$ is the dispersion in Stokes Q and U. We note that $\sigma$ includes both the observational uncertainties and the possible intrinsic dispersion within SDC13 due to geometrical depolarization or a variety of dust properties.

\begin{figure}
\includegraphics[width=\columnwidth]{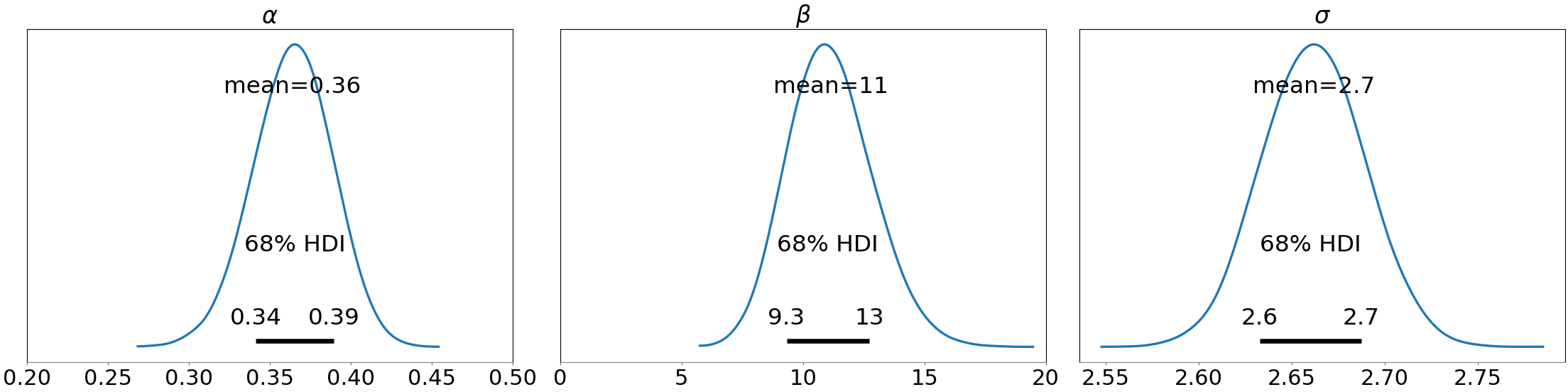}
\caption{Posterior distributions from Bayesian analysis for $I-P$ relation. The black lines delineate the 68\% ($1\sigma$) highest density interval (HDI). $\alpha$ is constrained to be significantly less than 1 with $0.36\substack{+0.03 \\ -0.02}$, suggesting that aligned dust grains are present within SDC13. }\label{fig:IPpost}
\end{figure}

The non-debiased polarization data are used in the Bayesian analysis, because the Ricean noise is well accounted for in this model. Only the pixels with higher uncertainties are excluded using the criteria $\sigma_I < 5$ m\jyb\ and $I/\sigma_I>3$. Note that a number of points with $P>10\%$ are still included in our samples, because these noisy-dominated points are important for the Bayesian analysis to model the noise component. The computed posterior distributions of the Bayesian analysis are shown in \autoref{fig:IPpost}. $\alpha$ is constrained to be $0.36\substack{+0.03 \\ -0.02}$ which is significantly smaller than 1. This suggests that aligned dust grains are present within SDC13, and hence the observed polarization patterns likely trace the magnetic field morphology. However, the observed distribution of polarization fraction appears to be more asymmetric than an ideal Ricean distribution, and thus a number of samples are located above the 98\% confidence region (CR). One possible origin of these high polarization fraction data are shock-compressed regions, where the magnetic field energy density is enhanced, and thus geometrical depolarization is suppressed. In addition, mechanical alignment torques originating from grains drifting through shocks are proposed to further align the dust grains \citep{ho18}, although more observational evidence is still needed to test this further.
Finally, other possibilities such as variations in dust grain properties might also play a role. 

\section{Histograms for all parameter pairs for individual regions in SDC13}\label{sec:allpair}
\begin{figure*}
\includegraphics[width=\textwidth]{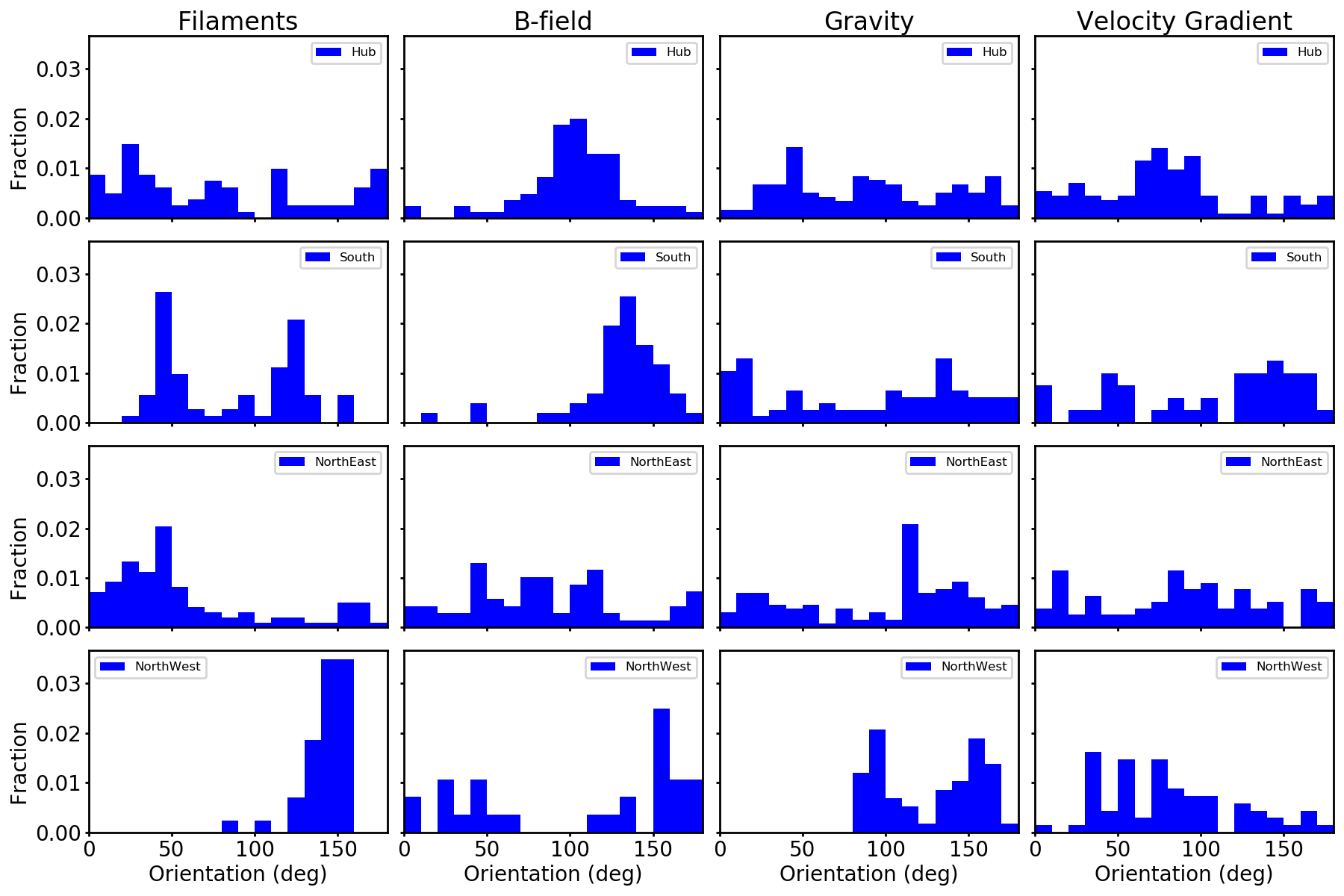}
\caption{Histograms of filament, magnetic field, local gravity, and local velocity gradient directions/orientations (columns) in the four SDC13 regions (hub, southern filament, northeastern filament, and northwestern filament; top to bottom row).}\label{fig:hist_region}
\end{figure*}

\begin{figure*}
\includegraphics[width=\textwidth]{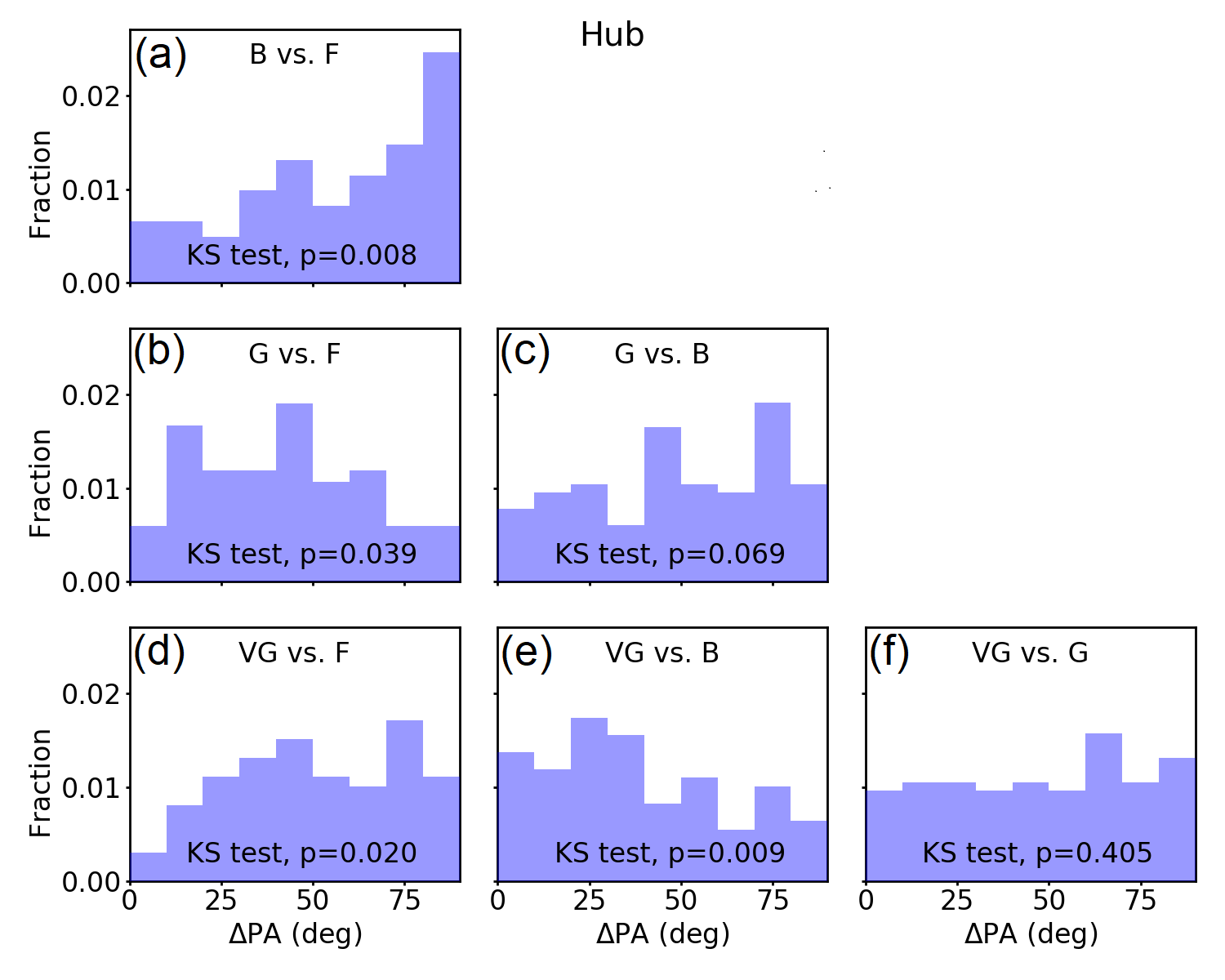}
\caption{Histograms of pairwise relative orientations among filaments, magnetic field, local gravity, and local velocity gradients in the central hub region.}\label{fig:hist_hub}
\end{figure*}

\begin{figure*}
\includegraphics[width=\textwidth]{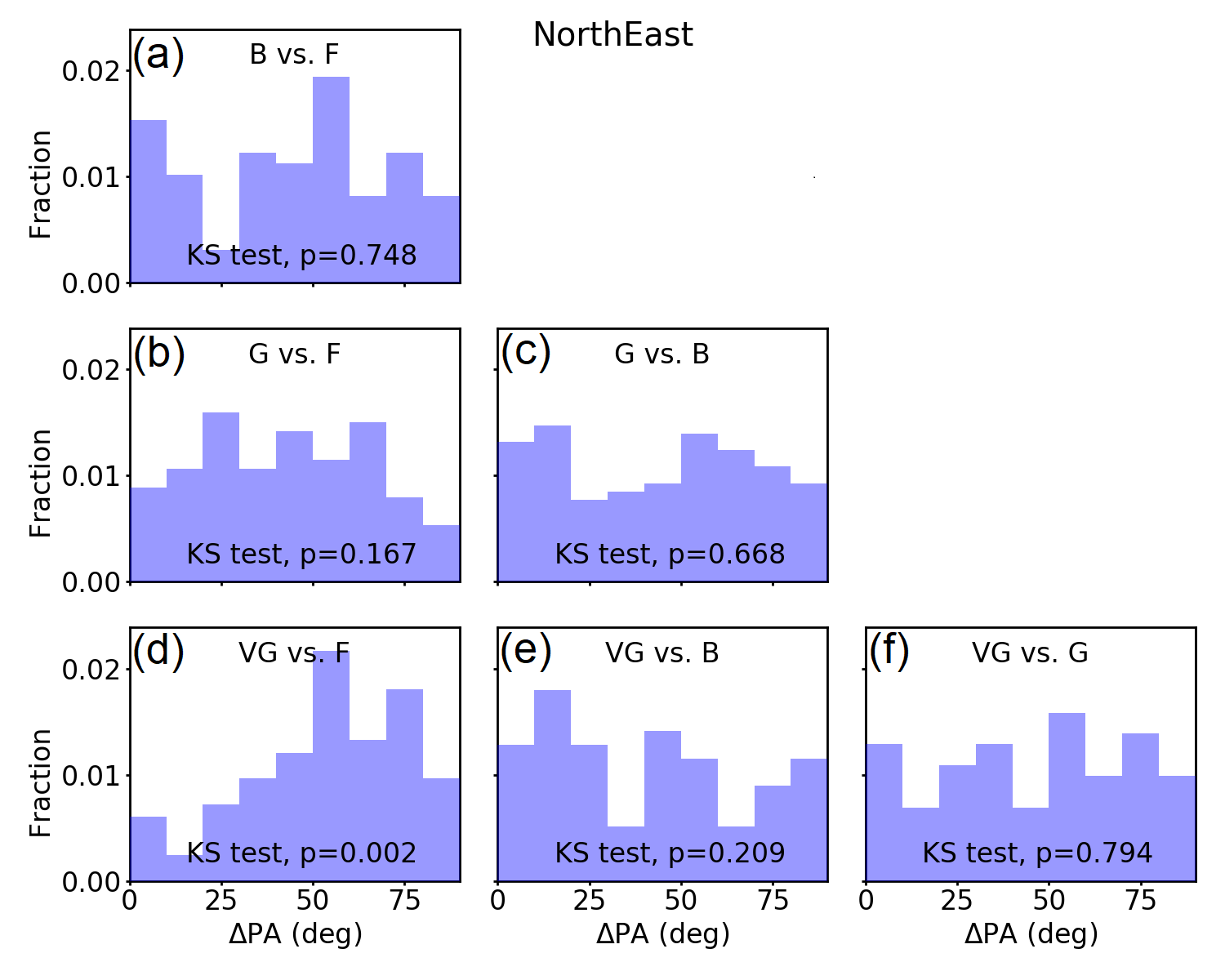}
\caption{Histograms of relative orientations among filaments, magnetic field, local gravity, and local velocity gradients in the northeastern filament region.}\label{fig:hist_northeast}
\end{figure*}

\begin{figure*}
\includegraphics[width=\textwidth]{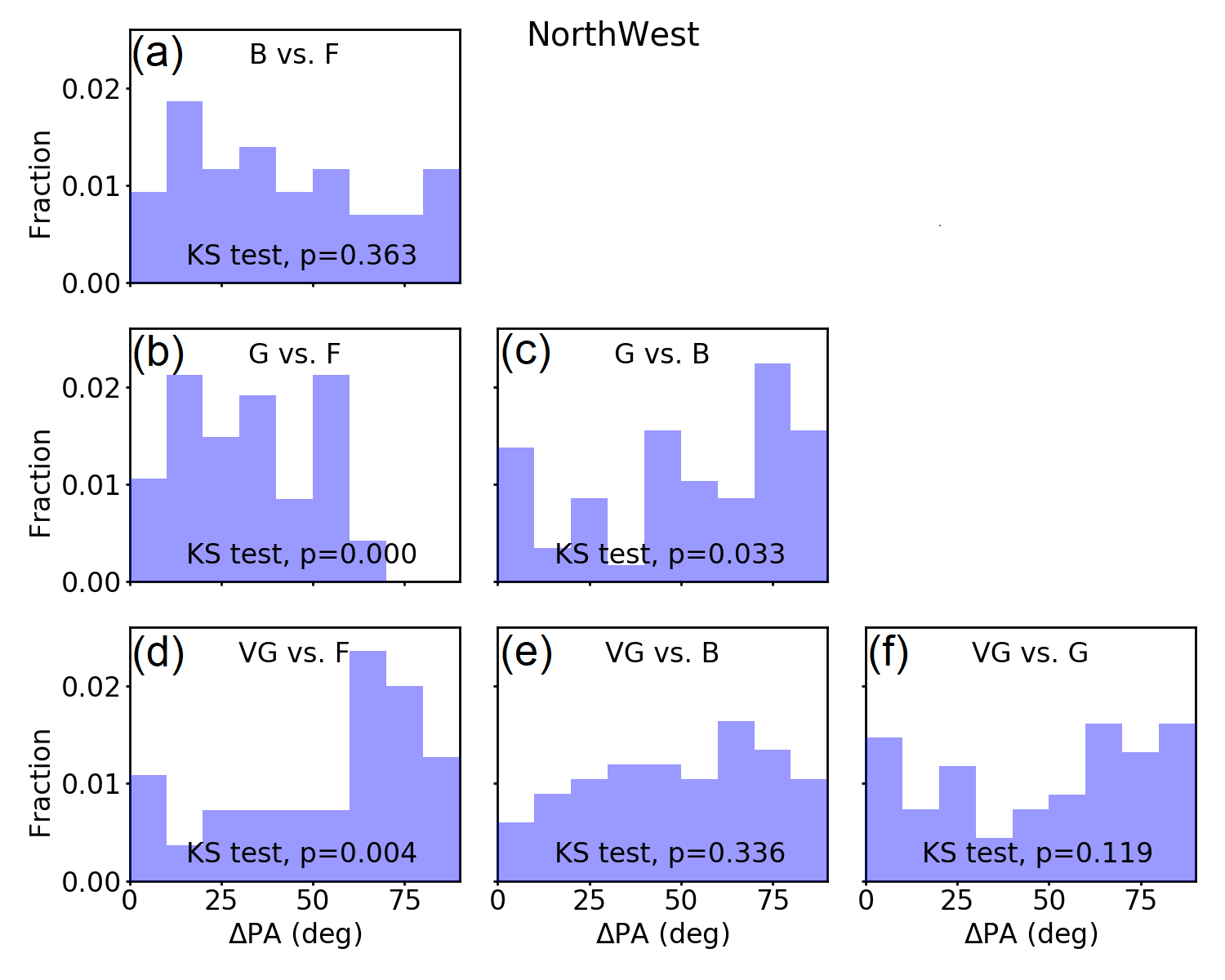}
\caption{Histograms of relative orientations among filaments, magnetic field, local gravity, and local velocity gradients in the northwestern filament region.}\label{fig:hist_northwest}
\end{figure*}

\begin{figure*}
\includegraphics[width=\textwidth]{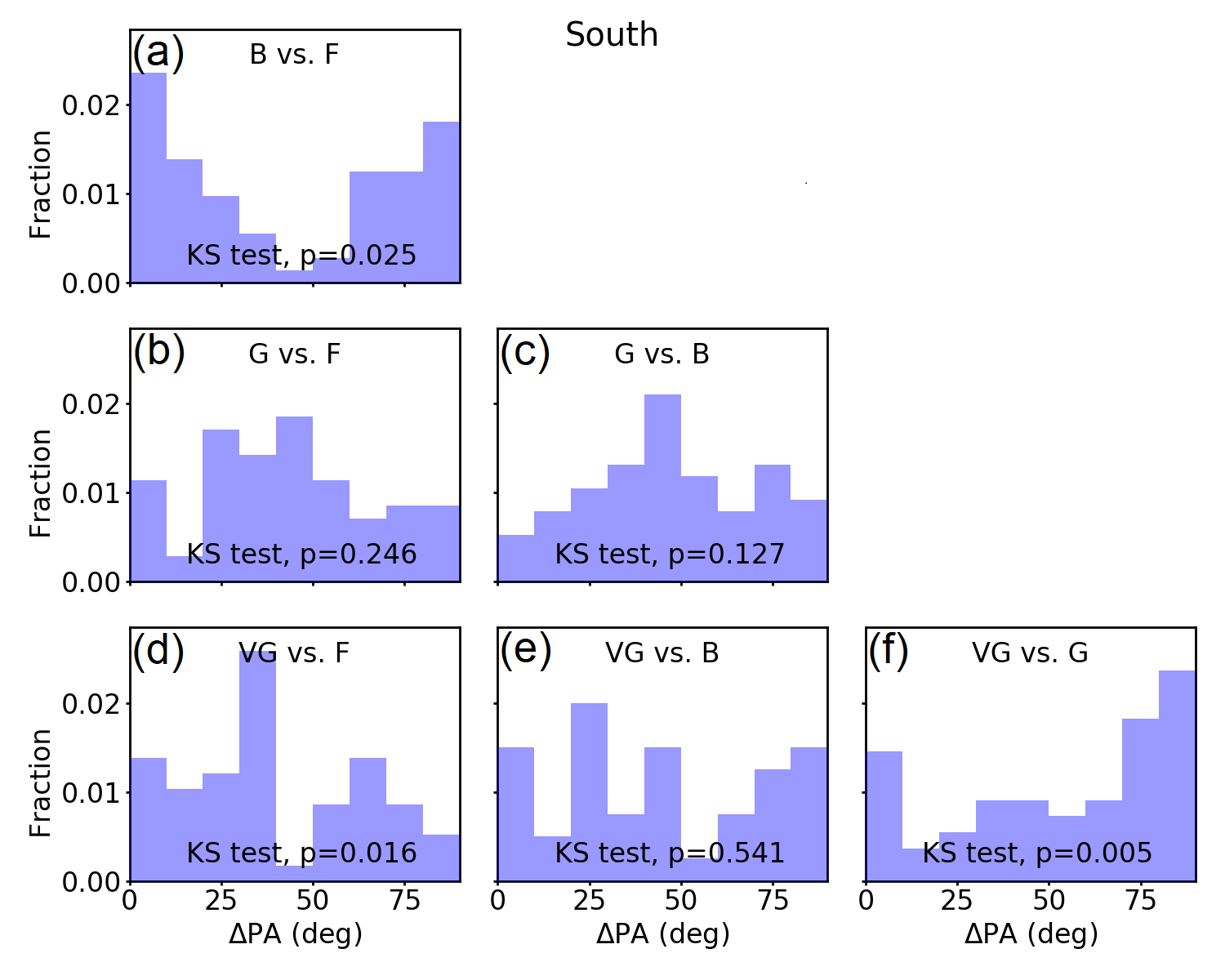}
\caption{Histograms of relative orientations among filaments, magnetic field, local gravity, and local velocity gradients in the southern filament region.}\label{fig:hist_south}
\end{figure*}

\begin{figure*}
\includegraphics[width=\textwidth]{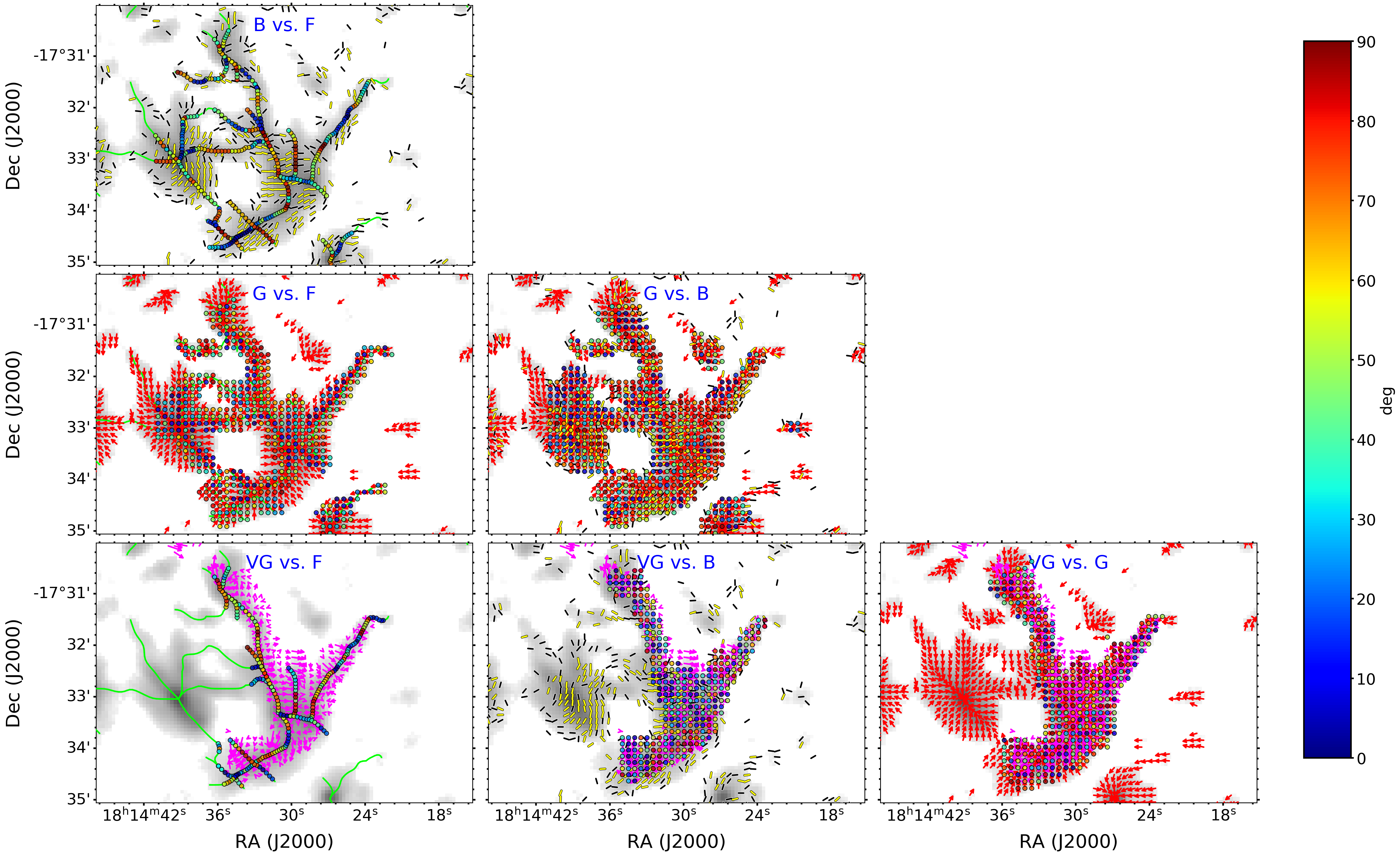}
\caption{Relative orientation maps for all pairwise comparisons between filament, magnetic field, gravity, and velocity gradient, overlaid on the JCMT 850 $\mu$m intensity. The color of a point indicates the relative orientation of a pair at each location. }\label{fig:dpa_map}
\end{figure*}

This appendix shows the complete histograms of the pairwise relative orientations for the individual regions in SDC13 (as labelled in \autoref{fig:regions}), 
based on the histograms for the grouped regions in \autoref{fig:hist_region}.
\autoref{fig:hist_hub},  \autoref{fig:hist_northeast},  \autoref{fig:hist_northwest}, and  \autoref{fig:hist_south} are the results for the hub, filament NE, filament NW, and filament S regions, respectively. 
\autoref{fig:dpa_map} is the map probing the spatial distribution of the relative orientation of all the pairs. Possible tendencies for each region are summarized in \autoref{tab:region_pair}.
Since these smaller subsets of data are likely affected by insufficient and incomplete coverage and resolution, they are not further discussed.

\begin{deluxetable*}{ccccc}
\tablecaption{Possible Tendencies of Physical Parameters for Separate Regions.\label{tab:region_pair}}
\renewcommand{\thetable}{\arabic{table}}
\tablenum{3}
\tablehead{\colhead{Pairs} & \colhead{Hub} & \colhead{Filament NE} & \colhead{Filament NW} & \colhead{Filament S} }
\startdata
\hline
B vs. F & $\perp(0.008)$ & ... & ... & $0\degr, 90\degr(0.025)$   \\
G vs. F & $25-60\degr(0.04)$ & ... & $\parallel(<0.001)$ & ...  \\
G vs. B & ... & ... & $\perp(<0.033)$ & ...  \\
VG vs. F & $\perp(0.02)$ & $\perp(0.04)$ & $\perp(0.004)$ &  $0\degr, 60\degr(0.016)$ \\
VG vs. B & $\perp(0.009)$ & ... & ... & ... \\
VG vs. G & ... & ... & ... & $\perp(0.005)$  \\
\enddata
\tablecomments{
Listed are p-values from KS-tests (in parentheses) only for pairs where p$<$0.05, i.e., a larger-than 95\% probability for a distribution to be different from random. All pairwise distributions for all separate regions are in \autoref{fig:hist_hub} to \autoref{fig:hist_south}.
Possible ranges and trends for relative orientations are noted ($\perp$: perpendicular; $\parallel$: parallel).
}
{\addtocounter{table}{-1}}
\end{deluxetable*}

\section{Nearby Compact Clouds}\label{sec:compact_clouds}
This appendix briefly describes the surrounding compact clouds detected by POL-2 and their possible connections with the large-scale environment. (The labeling refers to \autoref{fig:Bmap_all}.)

\subsection{SDC13.190-0.105}
This compact cloud is associated with bright 8--70 $\mu$m emission \citep{wi18}. It is possibly  more evolved than SDC13. Our POL-2 polarization data reveal a clear ``U-shape'' magnetic field pattern, pointing toward SDC13 following the large-scale magnetic field traced by \textit{PLANCK} in \autoref{fig:PLANCK}. This magnetic field morphology is consistent with the expectation of magnetic fields being dragged by accretion flows \citep{go18}. In addition, this compact cloud is part of a bigger complex, the giant molecular cloud (GMC) SDG013.222+0.0076, which is connecting SDC13.198-0.135, SDC13.190-0.105, and SDC13 following the large-scale magnetic field. All this together favors a scenario where this cloud is formed within a large-scale flow.

\subsection{SDC13.198-0.135}
This cloud is likely another fragment within the bigger GMC SDG013.222+0.0076. Unlike the ``U-shape'' magnetic field detected in SDC13.190-0.105, our POL-2 data reveal a toroidal-dominated magnetic field morphology. This morphology is consistent with the prediction of magnetic fields dragged by accretion flows in a late evolutionary stage \citep{go18}, where fragments in a large-scale flow are locally collapsing and the original ``U-shape'' magnetic fields are further twisted by rotating motions.

\subsection{SDC13.121-0.091}
This cloud is located at the conjunction between the GMC SDG012.840-0.2041 and the GMC SDG013.098-0.0821, with its major axis parallel to the conjunction boundary. The POL-2 magnetic field is aligned with the major axis of this compact cloud, but perpendicular to the large-scale
\textit{PLANCK} magnetic field in \autoref{fig:PLANCK}. This morphology suggests that this compact cloud is formed from the collision of the two GMCs, and thus morphology and magnetic field of this cloud are aligned with the compression layer.

\subsection{SDC13.123-0.157}
This system is composed of three dense clouds, distributed along the large-scale \textit{PLANCK} magnetic field in \autoref{fig:PLANCK}, and it is likely part of the GMC SDG013.222+0.0076. The local magnetic field within this system appears less clearly as a "U-shape" morphology, and it is perpendicular to the large-scale magnetic field and the cloud's major axis. This is consistent with a fragment formed within an accretion flow, similar to SDC13.190-0.105.

\subsection{SDC13.246-0.081}
This cloud is part of the GMC SDG012.840-0.2041. Both the \textrm{C$^{18}$O} and \nh3 velocity maps show that this compact cloud is likely connecting to the filament NE in SDC13. The local magnetic within this cloud is roughly aligned with the large-scale magnetic field along the filament, similar to the filament NE in SDC13. Hence, this cloud might be fragmented from filament NE.

\subsection{SDC13.225-0.004 and SDC13.177+0.017}
A series of compact clouds is distributed along the large-scale \textit{PLANCK} magnetic field in \autoref{fig:PLANCK}, as part of the GMC SDG013.222+0.0076. The local magnetic fields within these compact clouds are typically parallel or perpendicular to the large-scale magnetic field and the major axes of these clouds. This suggests that they might have fragmented from the GMC under the regulation of the large-scale magnetic field. However, we note that the overall structure of this system might be beyond the maximum recoverable scale of POL-2, and thus the small-scale magnetic field might not be adequately enough probed.

\facilities{JCMT,IRAM:30m}
\software{Aplpy \citep{aplpy2012,aplpy2019}, Astropy \citep{astropy2013,astropy2018}, DisPerSE \citep{so11}, FilChap \citep{su19}, GILDAS/CLASS \citep{pe05,gi13}, NumPy \citep{numpy}, SciPy \citep{scipy}, Smurf \citep{be05,ch13}, Starlink \citep{cu14}}

\bibliography{main}{}
\bibliographystyle{aasjournal}

\end{document}